\documentclass[12pt,preprint]{aastex}

\shorttitle{Cepheid Light Curve Reconstruction}
\shortauthors{Ngeow, Kanbur \& Nikolaev}

\begin{document}

\title{Reconstructing Cepheid Light Curve with Fourier Techniques I: \\
The Fourier Expansion and Interrelations.}

\author{Chow-Choong Ngeow and Shashi M. Kanbur}
\affil{Astronomy Department, University of Massachusetts,\\
Amherst, MA  01003}
\email{ngeow@nova.astro.umass.edu}

\author{Sergei Nikolaev}
\affil{Institute for Geophysics and Planetary Physics, \\ Lawrence Livermore National Laboratory,
Livermore, CA 94550}

\author{Nial R. Tanvir}
\affil{Department of Physical Science, University of Hertfordshire,\\ 
College Lane, Hatfield, AL10 9AB, UK}

\and
\author{Martin A. Hendry}
\affil{Department of Physics and Astronomy, University of Glasgow, \\
Glasgow G12 8QQ, UK}

\begin{abstract}

Fourier decomposition is a well established technique used in stellar pulsation. However the quality of
reconstructed light curves using this method is reduced when the observed data have uneven phase coverage.
We use simulated annealing techniques together with Fourier decomposition to improve the
quality of the Fourier decomposition for many OGLE LMC fundamental mode Cepheids. This method restricts
the range that Fourier amplitudes can take. The ranges are specified by well sampled Cepheids in the Galaxy and
Magellanic Clouds.

We also apply this method to reconstructing Cepheid light curves observed by the HST. These typically consist of
12 V band and 4 I band points. We employ a direct Fourier fit to the 12 V band points using the simulated
annealing method mentioned above and explicitly derive and use Fourier interrelations to reconstruct the
I band light curve. We discuss advantages and drawbacks of this method when applied to HST Cepheid data over existing
template methods. Application of this method to reconstruct the light curves of Cepheids observed in NGC 4258 shows that 
the derived Cepheid distance ($\mu_0=29.38\pm0.06$, random error) is consistent with its
geometrical distance ($\mu_0=29.28\pm0.09$) derived from observations of water maser.

\end{abstract}

\keywords{Cepheids---methods:data analysis---distance scale}

\section{Introduction}
 
    The study of Cepheid light curves has two major applications: (a) distance determinations to nearby galaxies via the well calibrated period-luminosity (PL) relations (for example, see \citet{mad85,fea87,mad91,sah00c}); (b) the understanding of the pulsational behavior of Cepheids by comparing the observed light curves to the theoretical counterparts \citep{sim83,buc90}. Therefore, reconstructing the Cepheid light curves from the observed photometric data is important regarding these two aspects. 

    Because Cepheid light curves are periodic, the data points from well observed Cepheids can be described by Fourier expansion. This technique was first introduced by Schaltenbrand \& Tammann in 1971, and developed by \citet{sim81} to study the structural properties of Cepheid light curves. In general, the $n^{th}$-order Fourier expansion has the following form:

     \begin{eqnarray}	
     m(t) = A_0 + \sum_{j=1}^{n}A_j\cos(j\omega t + \phi_j),
     \end{eqnarray}

    where $\omega=2\pi /P$ is the frequency and $P$ is the period of Cepheid. The $A_0$ term is the mean value of the light curve; the $A_j$ and $\phi_j$ are the Fourier amplitudes and phases for $j^{th}$ order, respectively. In this paper we
develop the use of simulated annealing techniques to improve the quality of
light curve reconstruction in nearby galaxies such as the LMC. We also apply these
techniques in the light curve reconstruction of Cepheids observed by Hubble Space Telescope (HST) which is necessary for distance determinations. In particular, we use Fourier expansion, as given by equation (1) but with modifications, to reconstruct the V band light curves and Fourier interrelations to reconstruct the I band light curves. These Fourier techniques have been established in stellar pulsation studies. We also present some potential advantages and drawbacks of the application of our method to this problem. 

     Section 2 \& 4 will describe the Fourier expansion and interrelations in details, respectively. Section 3 briefly describe the application of these Fourier techniques to HST data. The error analysis of the light curve reconstruction procedures will be presented in Section 5, following by the discussion and conclusion in Section 6. In appendix, we show another Fourier method, the Fourier intrarelations, that can be applied to reconstruct the Cepheid light curves. However, this method is less preferable than Fourier interrelations.

\section{Fourier Expansion}

    The form of Fourier expansion is given by equation (1), which has been extensively used in studying the light curves of pulsating stars (see \citet{ant87} and reference therein). Some applications of equation (1) in the study of the Cepheids can be found, for example, in \citet{sim81,sim83,sim85,ant86,and87,buc94,por94,hin97}. Obviously, equation (1) has $2n+1$ unknown parameters which require at least the same number of data points to solve for these parameters (in this paper we assume all of the periods for the Cepheids are given in the literature because they can be determined with other methods). Most of the observations on Galactic (eg., \citet{sch71,sim81,mof85,fer98}) and Magellanic Clouds Cepheids (eg., \citet{and87,mof98,uda99a,uda99b}) include large number of epochs that permit the higher order of Fourier expansion up to $n=8$. In contrast, there are typically about 12 V band epochs in the HST observations of extra-galactic Cepheids (see Section 3), which permit a $4^{th}$ order Fourier expansion. The $4^{th}$ order fit is a compromise between the goodness of fit and available data: higher order fits may not result in a better fit and lower order fits may not capture the structural properties of the light curves. Since there are only 4 I band data points in HST observations (see Section 3), the $4^{th}$ order Fourier expansion cannot be applied.

    Since the period is known from the literature, we can fold the time observation into phase as: $\Phi(t)=(t-t_0)/P - int[ (t-t_0)/P ]$, with $t_0$ being a common starting epoch. The value of $\Phi$ is from zero to one, corresponding to a full cycle of pulsation. Hence equation (1) can be written as \citep{sch71}:

    \begin{eqnarray}
      m(t) = A_0 + \sum_{j=1}^{4}A_j\cos[2\pi j\Phi(t) + \phi_j]
    \end{eqnarray}
      
\begin{figure}
      \plotone{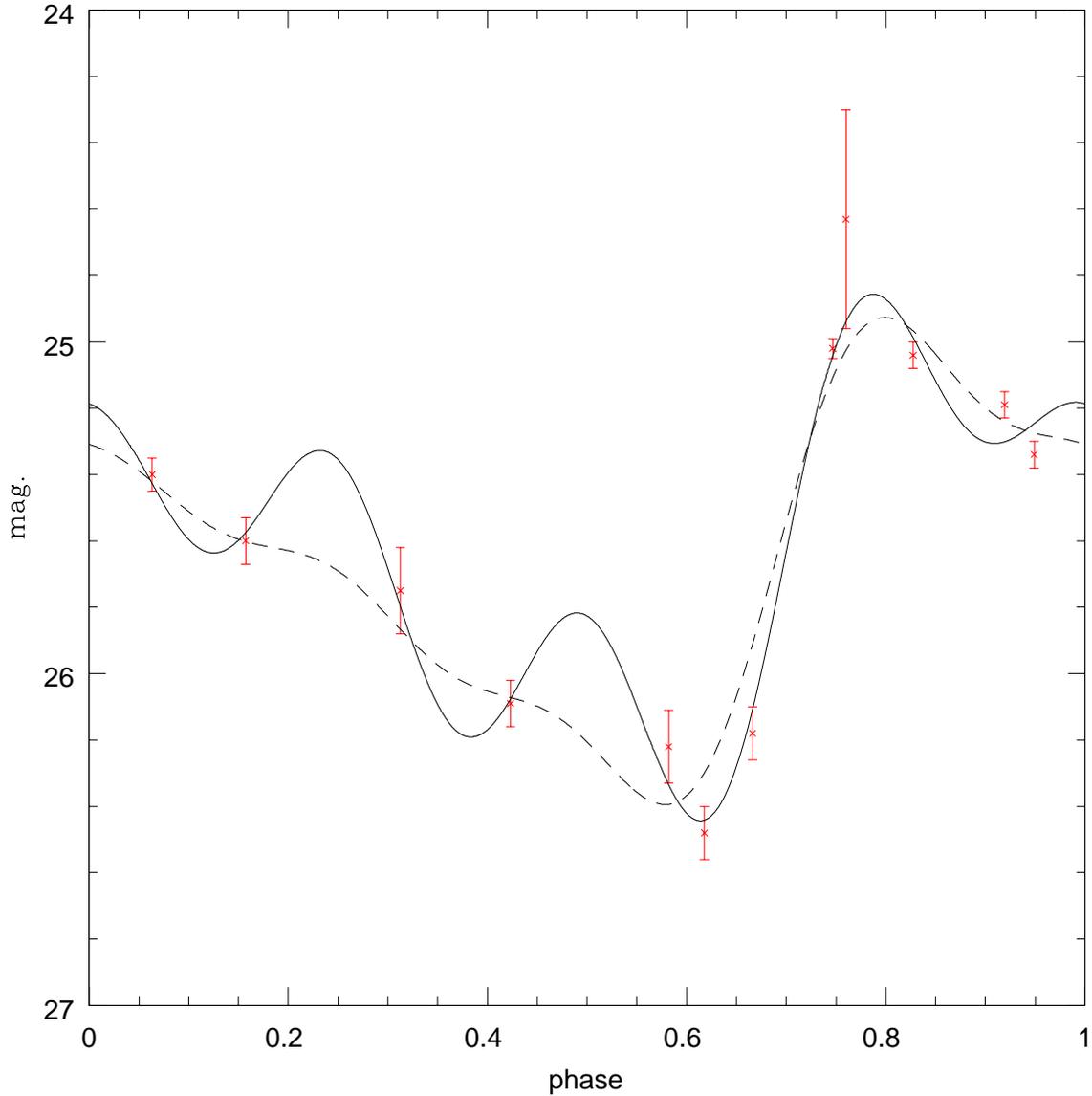}
      \caption{An example of the reconstructed light curve to a Cepheid in NGC 1326A. The solid curve is the reconstructed light curve from the least-square solution. It is clear that the least-square solution for this Cepheid does not provide a satisfactory fit to the data. Rather than solving for the parameters in equation (2), we fit the parameters in equation (2) with appropriate choice for the ranges (as given in Section 2.2) for Fourier amplitudes, $A_j$. The resulting light curve is drawn with a dashed curve, for comparison. The original data are indicated with error bars.\label{fig:fig1}}
    \end{figure}

    In general, one approach to solve the unknown parameters in equation (2) is by using the linear least-square method together with the available data. However, we find that in some cases the solution does not provide a good fit to the data and produces some numerical bumps in the light curves. Figure \ref{fig:fig1} illustrates this problem. The solid curve is the resulting light curve from the least-square solution to a Cepheid in NGC 1326A. The light curve does not fit to the data well and exhibits numerical bumps around phase $\sim0.2$ and $\sim0.5$. These numerical bumps are not physically associated with the true light curves, but result from the sparsely sampled data points. Instead of \textit{solving} for the unknown parameters, we specify the range of each parameter in equation (2) and \textit{fit} the data to obtain the best-fit values for each parameter. All the parameters are fit simultaneously with the simulated annealing method (for example, see \citet{pre92}) to minimize the corresponding $\chi^2$ value: $\chi^2=\sum [(m_{obs}-m_{fit})/\sigma_{obs}]^2$. The simulated annealing method will search the best fit values for a group of parameters within a reasonable amount of computational time. The dashed curve in Figure \ref{fig:fig1} shows that by fitting the parameters in equation (2) to the observed data with an appropriate choice for the ranges (given in Section 2.2 for Fourier amplitudes, $A_j$), the reconstructed light curve has been greatly improved. In the following subsections, we demonstrate how the choice of the ranges for Fourier parameters will affect the reconstructed light curves. 

\subsection{The Ranges of the Fourier Parameters}

     In order to apply the simulated annealing method to fit the available data points, certain ranges of the fitted parameters have to be imposed. The range for the mean magnitude ($A_0$) term in equation (2) is trivial (say, $\pm 5mag$ from the mean magnitude of the data points), and the range for the phases ($\phi_j$) is between $0$ and $2\pi$. For the Fourier amplitudes, the following ranges are \textit{initially} used to fit the Fourier expansion: 

	\[ A_1:0.0 - 0.7;\  A_2:0.0 - 0.5;\ A_3:0.0 - 0.5;\ A_4:0.0 - 0.5 \] 
     
\begin{figure}
   \plottwo{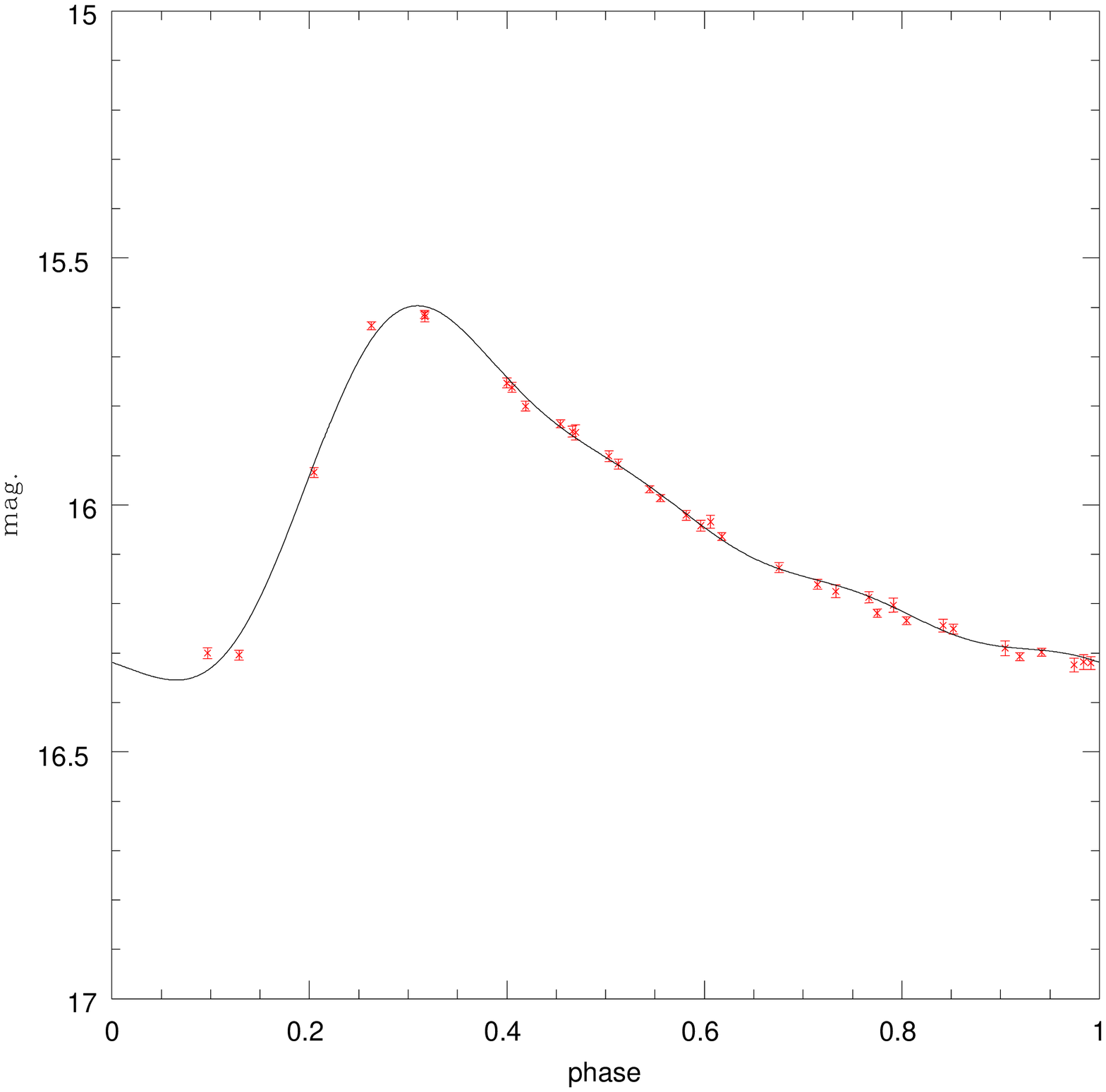}{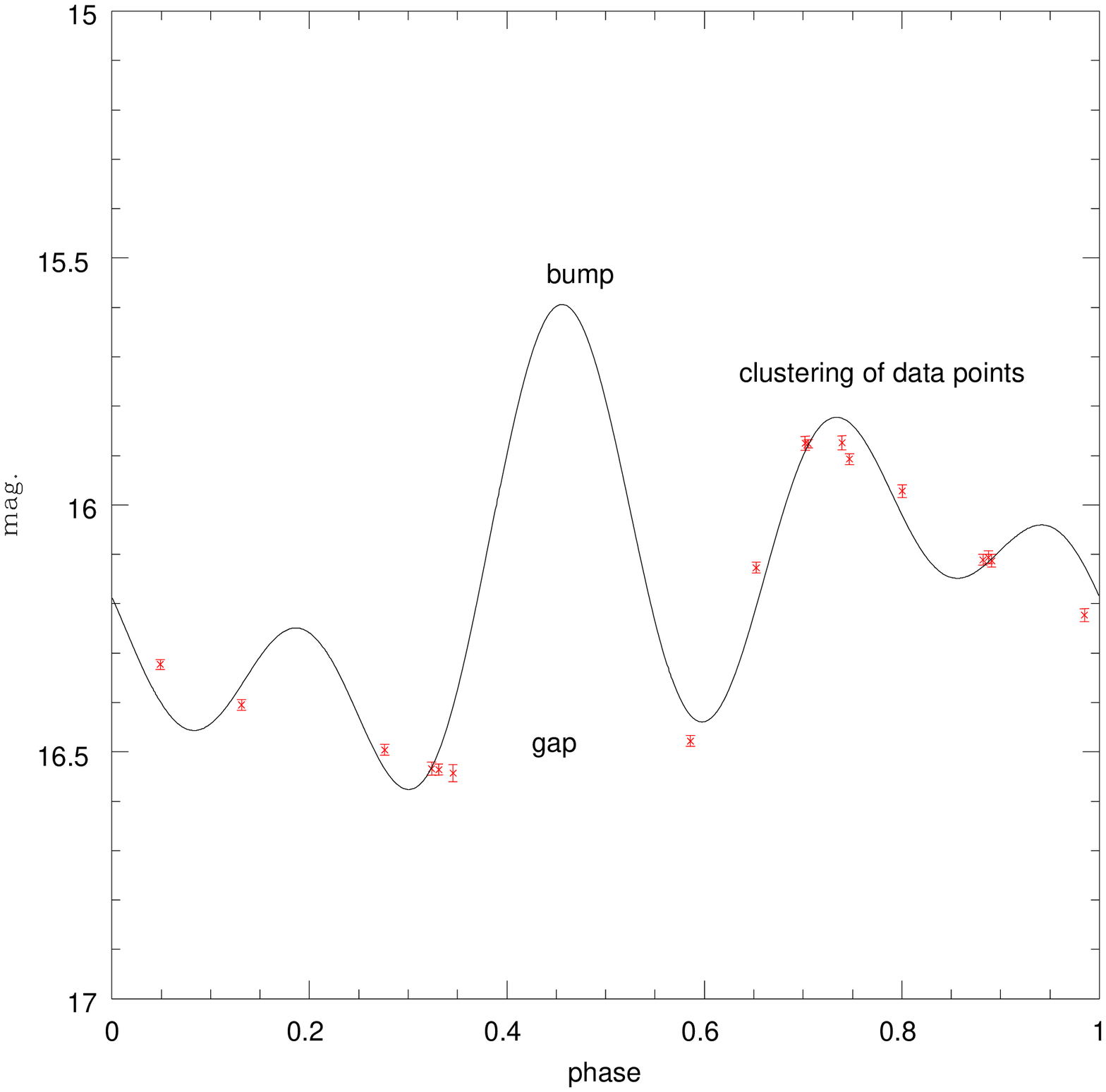}  
    \caption{Two typical examples for the constructed light curves (in V band) by using the simulated annealing method for LMC Cepheids with similar periods. The ranges for the Fourier phases that are used in fitting the data points are from $0$ to $2\pi$ and the ranges for the Fourier amplitudes are: $A_1:0-0.7;\ A_2, A_3, A_4:0-0.5$. These choice of ranges are relatively large, as demonstrated in Section 2.1. \textit{Left}: (a) Well constructed light curve. The data points are distributed relatively uniform in phase. This LMC Cepheid has $\log(P)=0.4576$. \textit{Right}: (b) Poorly constructed light curve due to the bad phase coverage, and exhibits a numerical bump at phase $\sim0.5$. This LMC Cepheid has $\log(P)=0.4580$.\label{fig:fig2}}
    \end{figure}

     These initial ranges are selected based on the guess that some Cepheids may have large Fourier amplitudes, but \citet{sim81} show the Fourier amplitudes are generally smaller than 0.4. Therefore we pick the above ranges as a starting point. By fitting the data with the ranges given above, most of the light curves can be reconstructed fairly well with good agreement to the data points. Figure \ref{fig:fig2}(a) shows an example of the reconstructed light curve for a Cepheid in the LMC with this initial choice for the ranges of Fourier amplitudes. However, these initial ranges also produce some poorly reconstructed light curves, due to the bad phased coverage of the data points, as illustrated in Figure \ref{fig:fig2}(b). Comparing the two cases in Figure \ref{fig:fig2} (which have similar periods), it is clear that some of the Fourier amplitudes for poorly reconstructed light curves are relatively large, indicating that the choice of the initial ranges is inappropriate. 

     The quality of the reconstructed light curve can be improved by further narrowing the ranges of the Fourier amplitudes when fitting equation (2) to the data points. In order to determine the proper ranges for the Fourier amplitudes, we use the OGLE LMC data (Optical Gravitational Lensing Experiment, \citet{uda99a}) which is obtained from the OGLE home-page\footnote{\url{http://bulge.princeton.edu/$\sim$ogle/}}. There are 771 fundamental mode Cepheids in the data, based on the judgment made by OGLE team \citep{uda99a}. The number of epochs in V and I band are $12-50$ and $>80$, respectively. The large numbers of epochs are sufficient to construct the light curves for individual Cepheids by using equation (2). However, the 10 Cepheids which do not have V band data and the Cepheids which are not fundamental mode Cepheids were removed from the final sample. Because the Fourier amplitudes are independent of distance, we also include a ``calibrating set'' of Cepheids \citep{hen99}, which comprise mostly of Galactic and some LMC/SMC Cepheids. In addition, most of the Cepheids observed by OGLE team have periods shorter than 10 days, with peak around $0.5<\log(P)<0.7$, and terminated at $\log(P)=1.5$. In contrast, the ``calibrating set'' Cepheids occupy a wider range of period, from $\log(P)\sim 0.2$ to $\log(P)\sim 2.0$. Therefore, including the ``calibrating set'' Cepheids can be used to determine the ranges of Fourier amplitudes beyond $\log(P)=1.5$. There are about 118 Cepheids in this sample. Since I band data for OGLE LMC has a large number of epochs, we mainly impose ranges for the Fourier amplitudes in V band.  

     \textbf{The V Band Fourier Amplitudes:} After constructing the light curves of OGLE LMC Cepheids with initial (and relatively large) Fourier amplitudes, we divided the V band light curves into two groups by visual inspection: those with well constructed light curves (as in Figure \ref{fig:fig2}(a)) and those with poorly constructed light curves (as in Figure \ref{fig:fig2}(b)). The distributions of the Fourier amplitudes as a function of $\log(P)$ are plotted in Figure \ref{fig:fig3}. This shows the positions of Fourier amplitudes for well constructed light curves (in crosses) and poorly constructed light curves (in triangles), as a function of period. As can be seen from the figure, the distributions of the Fourier amplitudes for well constructed light curves occupy a smaller range than the poorly constructed light curves, which have relatively larger amplitudes especially for higher order harmonics. The V band light curves are further trimmed to contain only the really well constructed light curves, exemplified by Figure \ref{fig:fig2}(a). We then re-plot the distributions in Figure \ref{fig:fig4}, which shows evidence of well defined ranges for the V band Fourier amplitudes.   
   
     \textbf{The I Band Fourier Amplitudes:} Since the number of epochs for most of the OGLE LMC Cepheids in the I band is more than 80, the constructed light curve are expected to be well behaved and the Fourier amplitudes are assumed to fall within the appropriate ranges. The plots of the Fourier amplitudes are shown in Figure \ref{fig:fig5}, with some suspected outliers marked in triangles and labelled. We also plot out some of the Fourier light curves for these outliers in Figure \ref{fig:fig6}. As seen from the figure, some of them have acceptable light curves, with slightly larger amplitudes when compared to the Fourier light curves from Cepheids with similar periods. The exception is $C686$, which exhibits a tip around phase 0.1 from the original data. The simple $4^{th}$ order Fourier expansion cannot reproduce this tip and hence the constructed light curves do not show good agreement to the data. Nevertheless, the difference in amplitudes for the outliers and the majority of the data points are very small, because the I band amplitudes for Cepheids are generally smaller \citep{fre88}.  

\subsection{The Adopted Ranges for Fourier Amplitudes} 

     From Figure \ref{fig:fig3}-\ref{fig:fig5}, it is clear that the Fourier amplitudes occupy certain ranges in the amplitude vs. $\log(P)$ plots. Therefore we can determine the appropriate ranges of Fourier amplitudes for a given period from these figures. The adopted ranges for the Fourier amplitudes in V and I band are presented in Table \ref{tab1}. Due to the fact that the OGLE LMC data only contains Cepheids with periods less than 32 days ($\log(P)<1.5$), the ranges for Cepheid with $\log(P)>1.5$ are determined from the ``calibrating set'', while the ranges for shorter period Cepheids are determined from both OGLE LMC and ``calibrating set'' data. In addition, the ranges of the Fourier amplitudes are set to be started from zero to account for possible low amplitude Cepheids. Note that the upper limit of the ranges given in Table \ref{tab1} is an approximation, since there is no exact upper limit for a given period. Experience shows that sometimes a slightly larger range of the Fourier amplitudes than the one given in Table \ref{tab1} can reconstruct the light curve better. An iterative process can be made, if necessary, to find the most suitable upper limits to reconstruct satisfactory light curves. The reduction in ranges can improve the quality of the Fourier fit, as some of the numerical bumps are removed in the reconstructed light curve, provided that the problem of bad phase coverage is not too severe (see details in Section 2.3). 
 
     The distributions of the Fourier amplitudes for OGLE LMC Cepheids and the ``calibrating set'' (or mostly Galactic) Cepheids appear to coincide, as seen in Figure \ref{fig:fig3}-\ref{fig:fig5}. This may imply that the ranges of the Fourier amplitudes depend weakly on metallicity. However, this conclusion is only based on the analysis of two galaxies, and may not reflect the assumption that metallicity can affect the distribution of Fourier amplitudes. \citet{van78} showed that the upper limits of B band amplitudes are different for Cepheids in Galaxy/M31, LMC and SMC. In addition, \citet{pac00} showed that the OGLE LMC Cepheids, in the period range of $1.1<\log(P)<1.4$, have larger amplitudes than OGLE SMC Cepheids with same period ranges. Because the V band amplitudes can be scaled from B and I band amplitudes \citep{fre88}, the different V band upper limits for Cepheids in different galaxies could also exist. Nevertheless, the ranges of Fourier amplitudes given in Table \ref{tab1} are assumed to cover the Fourier amplitudes for different metallicity environments, because these ranges go from zero to the upper limit that is slightly larger than the ranges defined by OGLE LMC and ``calibrating set'' Cepheids. The detailed analysis of the relationship between the Fourier amplitudes and the metallicity environments will be presented in future work.

\subsection{Results and Problems of the Fits} 

     Using the techniques of $4^{th}$ order Fourier expansion with the appropriate ranges for the Fourier amplitudes, some of the poorly reconstructed light curves can be improved and the results show good agreements to observational data. Figure \ref{fig:fig7} shows some examples of the improved reconstructed light curves when fitting equation (2) to the OGLE LMC V band data by narrowing the Fourier amplitudes as given in Table \ref{tab1}. As can be seen from the figure, some of the obvious bumps (or dips) in the original reconstructed light curves can be removed with this simple technique, resulting in smoother light curves and better fits to the data. The estimation of the mean magnitudes would be closer to the \textit{true} mean without the influence of the numerical bumps/dips. 

     In certain extreme cases that either the data points are heavily clustered in certain phases or there is a large gap between two adjacent data points, the Fourier expansion fails to construct a satisfactory light curve, regardless of the choice of the ranges for Fourier amplitudes. Figure \ref{fig:fig8} shows some examples of this case. By using either the initial ranges (dotted curves) or the narrower ranges (solid curves) to fit the Fourier amplitudes with observed data points, the reconstructed light curves are still not satisfactory due to the phase clustering of the original data. When using this technique to estimate means for the PL relations, these Cepheids should be treated with caution or rejected. In fact, the rejection of Cepheids with bad phase coverage has been practiced in the past (for example, see \citet{sim81,ant86}).  

\section{Application to HST Data}

Since Cepheids are primary distance indicators, and the Cepheid PL relations can be used to calibrate secondary distance indicators, such as the Tully-Fisher relation, Type Ia supernovae, fundamental plane and surface brightness fluctuations. This builds up the distance ladder and indicates the distance to more remote galaxies. Together with the measurements of the recession velocities of these more distant galaxies, the Hubble constant can be determined: this was the main goal of Hubble $H_0$ Key Project (\citet{fre01}, hereafter H0KP).
  
     The conventional way of applying the Cepheid PL relations is by measuring the mean magnitude of the Cepheid over the pulsational cycle for a given period. In recent years, a great deal of effort has been made to discover extra-galactic Cepheids, and determine the distance, by using the HST (as in H0KP and \citet{sah99,sah01a,sah01b}). In most of the HST observations, in order to optimize the phase coverage of Cepheids within the HST observational windows, the number of V band observations is chosen to be 12 \citep{fre94,ken95}. To correct for the extinction and reddening, I band observations are included (see, for example, HOKP and \citet{ken95}). However, only 4 (or 5) I band observations are needed, because the I band amplitudes are smaller than in the V band \citep{fre88,fre94,ken95}. Existing techniques to estimate the mean from the sparsely sampled HST data include: (1) taking the phased weighted intensity mean \citep{sah90} in V band and using an empirical relation developed by \citet{fre88} in I band (for example, see \citet{sil96}); (2) adopting template fitting procedures \citep{ste96,tan99}. 

      In the application of the methods developed in the previous sections to this problem, we use
we use Fourier expansion, as given by equation (1) but with modifications, to reconstruct the V band light curves and Fourier interrelations to reconstruct the I band light curves. Some results of applying these methods to determine the Cepheid distance in nearby galaxies are presented in Section 6.4.

\section{Fourier Interrelations}

    Because the I band observation of HST typically consist of 4 epochs, equation (2) cannot be applied to reconstruct the light curves. However, statistical relationships between the $j^{th}$ order Fourier coefficients for V and I band have been introduced by \citet{hen99} and by \citet{kan01}, known as the Fourier interrelations. We explicitly derive and present them here because of their applicability in stellar pulsation studies. These allow the reconstruction of I band light curves up to $4^{th}$ order. There also exist correlations of the first order Fourier amplitude ($A_1$) to subsequent higher order Fourier amplitudes in the same bands (for example, see \citet{ant87}), known as the Fourier intrarelations  (presented in Appendix), this approach is less preferable than Fourier interrelations. The Fourier interrelations are the linear relation of $j^{th}$ order Fourier amplitude and phases from V band to I band, and have the following form:

    \begin{eqnarray}
      A_j(I) & = &\alpha_j + \beta_j A_j(V),\ \ j=1,\ldots 4 \\       
      \phi_j(I) & = & \gamma_j + \eta_j \phi_j(V),\ \ j=1,\ldots 4 
    \end{eqnarray}
    
    The coefficients of the interrelations are determined from the ``calibrating set'', by minimizing the $\chi^2$ value of the fit. Fourier interrelations for fundamental mode Cepheids are presented in Table \ref{tab2}. The table also shows the rms of the straight-line fit to the data. The fits were derived taking full account of errors in both variables with the standard model fitting procedures (for example, see \citet{pre92} Chapter 15). The Fourier interrelations for equation (3) \& (4) are clearly shown in Figure \ref{fig:fig9} \& \ref{fig:fig10}, respectively. The standard errors of the Fourier coefficients, derived from inverting the Hessian matrix, are also shown in the figures. Despite the appearance of strong correlations in the figures, the relations have large $\chi^2$ per degree of freedom, i.e. very small statistical significance. Nevertheless, using interrelation to recover sparsely sampled light curves works quite well, as shown in Section 6. For the Galactic data in the ``calibrating set'', we display bump and non-bump Cepheids with solid and open squares respectively (due to the $P_2/P_0=0.5$ resonance around 10 days on Hertzsprung progression, the periods of bump Cepheids normally lie in between 8 and 14 days.). For the LMC/SMC data, we do not differentiate between bump and non-bump Cepheids. 

    Besides the Fourier interrelations among the ``calibrating set'' Cepheids, similar linear relations also exist for OGLE LMC and SMC Cepheids. The Cepheid data of OGLE LMC are as in Section 2, and the data of OGLE SMC \citep{uda99b} are downloaded from OGLE home-page. The number of the fundamental mode Cepheids, as judged by \citet{uda99b}, observed in SMC is $\sim1300$, with most of them in short period ranges. Table \ref{tab3} presents the best-fit results of Fourier interrelations to the OGLE LMC data, and the plots of the Fourier interrelations are given in Figure \ref{fig:fig11} \& \ref{fig:fig12} for Fourier amplitudes and phases, respectively. The preliminary results of the OGLE SMC Fourier interrelations are presented in Table \ref{tab4}, with the plots of Fourier amplitudes and phases in Figure \ref{fig:fig13} \& \ref{fig:fig14}, respectively. Again a clear linear relation exists with increasing scatter as we go to higher order parameters. There is little difference in the slopes of the best fit lines to that found for ``calibrating set'' Cepheids. Hence, the Fourier interrelations given in Table \ref{tab2} will be applied to reconstruct the I band light curves.  

    The procedures for reconstructing the I band light curves are similar to V band light curves. Instead of fitting all the parameters (the mean magnitude and the Fourier parameters), we use the Fourier interrelations to obtain the I band Fourier parameters from the V band fit. This gives the shapes of the I band light curves, and the mean I band magnitudes are found by using the observed I band data to minimize the corresponding $\chi^2$ values. 

    There is little difference between the Fourier interrelations for Cepheids in ``calibrating set'' and in OGLE LMC or SMC, as shown in Table \ref{tab2}, \ref{tab3} \& \ref{tab4}, respectively. The good agreement between them indicates that the {\em relative} changes in Fourier parameters are almost unaffected by metallicity. This does not mean, however, that the light curve shape is independent of metallicity. Rather, it means that change in metallicity affects the Fourier parameters in different bands in the same way. Therefore, given a well-sampled light curve in V band, one can estimate Fourier parameters in I band reasonably well regardless of the metallicity of the parent galaxy. This result is useful in reconstructing the light curves of extra-galactic Cepheids in a broad range of metallicity environments.    

     The empirical relation that the earlier papers of H0KP used to obtain the I band mean magnitude is based on the observational results of \citet{fre88}, as the amplitude ratio of I and V band is about 0.5. However, \citet{tan97} recommended use of an amplitude ratio of 0.6, because this ratio can improve the estimation of I band mean magnitude. We note that the slopes in the Fourier interrelations ($\beta$ term in Table \ref{tab2} \& \ref{tab3}) are approximately equal to 0.6, which is close to the Tanvir's value. This is the main reason why Fourier interrelation exists and work well in reconstructing the light curves (Section 6).

\section{Error Analysis for Light Curve Reconstruction}

     The Fourier techniques described in this paper are now applied to reconstruct the Cepheid light curves in nearby galaxies observed by HST, particularly for H0KP galaxies. Perhaps the most important question regarding the application of our work to this problem is the following: does a fourth order fit with 9 free parameters to 12 V band data points represent anything physically? In order to estimate the errors associated with the light curve reconstruction procedures presented in Section 2 \& 4, we performed an error analysis based on Monte-Carlo simulations. 

     Due to the large number of Cepheids, as well as the large number of observations in V and I bands, the simulations are performed primarily with OGLE LMC Cepheids that have good light curves (as in Figure \ref{fig:fig2}(a)) and represented by crosses in Figure \ref{fig:fig3}. These OGLE LMC Cepheid light curves, constructed by using all the available data points in both bands, are referred as the original light curves. We assume that when we fit a $4^{th}$ order Fourier expansion to these data, the resulting Fourier amplitudes are very close to their true values.   

    In order to mimic the published HST photometric data, we performed three simulations for the error analysis. The procedures for each simulation are as follows: 
	
	\begin{itemize}
		\item \textbf{Simulation 1}: First a Cepheid is picked randomly (with replacement) from the data set, then 12 points in the V band and 4 points in the I band were randomly selected without replacement. In addition to the OGLE LMC photometric errors, we add Gaussian noise of $\mu_{noise}=0.05mag$ and $\sigma_{noise}=0.10mag$ to these randomly selected data. 
		\item \textbf{Simulation 2}: Same as in Simulation 1, but with larger Gaussian noise of $\mu_{noise}=0.15mag$ and keep $\sigma_{noise}=0.10mag$.
		\item \textbf{Simulation 3}: We pick one Cepheid with large number of epochs ($N_v=34,\ N_I=188$), then 12 points in the V band and 4 points in the I band were randomly selected without replacement form this Cepheid. The additional Gaussian noise is same as in Simulation 1.
	\end{itemize}

    Then the data from each simulation are used to reconstruct the light curves with the procedures described in Section 2 \& 4, i.e. $4^{th}$ order Fourier fit in the V band and interrelations in I band. The randomly selected 12 V points could of course be uniformly distributed or concentrated around one particular phase point. Mean magnitudes and the Fourier amplitudes obtained from our reconstruction procedures for these simulated data are then compared to the mean magnitudes and Fourier amplitudes from the original light curves. This is repeated a large number of times to build up an error distribution. 
    
    We emphasize that we are simulating the photometric data published in the literature. It has been suggested by an anonymous referee that even this is not a real simulation of HST Cepheid data since we neglect the possibility of events such as "warm pixels" or cosmic rays. However we contend that such data points will either be rejected by the photometric reduction package used and, if not, the point sources responsible for these should not be used anyway. 

     After running $N=1000$ trials in each simulations, the error histograms for the mean magnitudes, as well as the Fourier amplitudes, in both V and I bands are constructed. Gaussian distributions with parameters of $\mu$ and $\sigma$ are then fitted to these error histograms, where $\mu$ represent the mean offsets between the simulated data and original data, and $\sigma$ represent the errors in either the means or the Fourier amplitudes. The results of the Gaussian fits from each simulations for V and I bands are presented in Table \ref{tab5} \& \ref{tab6}, respectively. The error histograms for the mean values, resulted from Simulation 1, are presented in Figure \ref{fig:fig15}, with abscissa to be the difference between the simulated means and original means. Similarly, the errors histogram for Fourier amplitudes in V and I bands are presented in Figure \ref{fig:fig16} \& \ref{fig:fig17}, respectively. The parameters from Gaussian fits are listed in the upper-left corners of these figures. We did not include the error histograms for Simulation 2 \& 3 in this paper, because they all look similar to the error histograms of Simulation 1 (Figure \ref{fig:fig15}, \ref{fig:fig16} \& \ref{fig:fig17}). 

     To check the simulations, we also plot the V band $R_{21}$(=$A_2/A_1$) parameters \citep{sim81}, resulting from Simulation 1, as function of $\log(P)$ in Figure \ref{fig:fig18}. The filled circles are the original data and the crosses are the simulated data. Those crosses with triangles are the simulated data when either $A_1$ or $A_2$ is about $2.0\sigma$ away from the original values. It can be seen from the figure that the simulated data do indeed trace the original data. In addition, the $P_2/P_0=0.5$ resonance around 10 days can be clearly seen from this figure.   
  
     From Table \ref{tab5} \& \ref{tab6}, it can be seen that the errors of the means and each Fourier amplitudes are consistent in all three simulations. As expected, Simulation 2 gave the largest errors because the Gaussian noise generated in this simulation is bigger that other two simulations. Simulation 3 has the smallest errors, because this simulation is performed for one star. The errors for Simulation 1 is in between these two cases, which is typical. In addition, Table \ref{tab5} \& \ref{tab6} shows that the mean offsets of the simulated data and original data are very small. This can be seen in  Figure \ref{fig:fig15}-\ref{fig:fig17}, as all histograms are almost centered at zero. Therefore, no bias introduced by the reconstruction procedure, because all four Fourier amplitudes are tightly clustered around the "true" values given by the original well sampled Cepheid.

    The simulations suggests that with typical HST Cepheid data, it is meaningful to fit a fourth order Fourier expansion and that
our direct Fourier fitting procedures for estimating $V$ and $I$ band means are not biased.

\section{Discussion and Conclusion}

     The light curve of a Cepheid can be reconstructed with the Fourier techniques described in this paper. For OGLE LMC Cepheid data, we have demonstrated that we can frequently improve the quality of the reconstructed light curve by restricting the range of Fourier amplitudes can take in the Fourier fit. These ranges are obtained from well sampled Cepheid light curves in the Galaxy and Magellanic Clouds.

     We have then applied these techniques to HST Cepheid data. Such data typically consist of 12 and 4 V and I band points respectively. In particular, Cepheids with sufficient observations permit a $4^{th}$ order Fourier expansion. However, the Cepheids in HST I band observations that have insufficient data points require the application of Fourier interrelations to reconstruct the I band light curves from the V band. In summary, the Fourier techniques for light curve reconstruction procedures, as applied to the HST data, include two main parts:

     \begin{itemize}
     \item  $4^{th}$ order Fourier expansion to 12 V band data points. This is reconstructing the light curves by a direct fit to the data points, with appropriate ranges for each of the Fourier parameters.
     \item  Fourier interrelation to 4 I band data points. This is using the linear relations of Fourier parameters between V and I bands to reconstruct the light curves. 
     \end{itemize}

     In general, these relatively simple techniques can reconstruct the Cepheid light curves quite well, given that the data points are relatively uniform and well sampled. These techniques serve as an alternative method to obtain the mean magnitude of Cepheids besides phase weighted intensity mean and template fitting procedures. However, the detailed comparisons between these methods are beyond the scope of this paper. Some examples of the reconstructed light curves for extra-galactic Cepheids (observed by H0KP) with these Fourier techniques are presented in Figure \ref{fig:fig19}, which show good agreements to the observed data points. 

     We also performed Monte-Carlo simulations to determine the errors associated with our light curves reconstruction procedures. The errors of the mean magnitudes and Fourier amplitudes are determined from the fits of Gaussian distribution to the error histograms. These Gaussian fits show that our procedure is unbiased. The errors are: 
	\begin{itemize}
	\item	V band: $\sigma_{mean}=0.034$; $\sigma_{A1}=0.039$; $\sigma_{A2}=0.033$; $\sigma_{A3}=0.028$ and $\sigma_{A4}=0.023$.
	\item	I band: $\sigma_{mean}=0.035$; $\sigma_{A1}=0.025$; $\sigma_{A2}=0.021$; $\sigma_{A3}=0.017$ and $\sigma_{A4}=0.014$. 
	\end{itemize} 

Could it be the case that the distribution of 12 points is so severe that a decent fit to the V band data is not possible?
Certainly and it could be that in this case a template type technique will produce a V and I band mean.
However, we again contend that in this case, it will probably be the case that this point source is not used in the
final analysis.

\subsection{The Correlations of Fourier Amplitudes}

     It has been suggested by anonymous referee that the higher order Fourier amplitudes shown in Figure \ref{fig:fig3} \& \ref{fig:fig5} are correlated with the first order Fourier amplitude ($A_1$) in the same bands (the Fourier intrarelations).
These have been examined, for example, by \citep{ant87} and
in the appendix. However, the Fourier intrarelations are less preferable than Fourier interrelations for reconstructing the Cepheid light curves. The reasons are discussed in following examples and in Appendix. 

     Here, we reexamine the results of \citet{ant87} with our data sets and compare them to the Fourier interrelations. We used all ``calibrating set'' Cepheids and pick only the ``good'' OGLE LMC Cepheids (crosses) from Figure \ref{fig:fig3}. Then we re-plot the $A_1-A_2$ Fourier intrarelations and the $A_1(V)-A_1(I)$ Fourier interrelation in Figure \ref{fig:fig20}. The ``Calibrating set'' Cepheids and OGLE LMC Cepheids are in left and right panels, respectively. In the figure, we distinguished Cepheids with different period ranges: crosses are for Cepheids with periods shorter than 8 days (short-period Cepheid); triangles are for Cepheids with period in between 8 to 14 days (bump Cepheid); and filled circles are for Cepheids with periods longer than 14 days (long-period Cepheid). Error bars for each data point are omitted in the figure for clarity. The top two panels in Figure \ref{fig:fig20} are Fourier intrarelations in V and I band respectively, and the bottom panel is the Fourier interrelations of $A_1(V)$ and $A_1(I)$. 

     It is clear from the figure that the scatter of $A_1-A_2$ intrarelations are larger than the scatter of $A_1(V)-A_1(I)$ interrelations. Furthermore, the short-period, bump and long-period Cepheids populate different regions in the plots of intrarelations, as is also seen in \citet{ant87}. In contrast, The tightness of correlations in the Fourier interrelations is clear, and less dependent on period distribution. These two properties make Fourier interrelations more applicable in reconstructing the light curves than Fourier intrarelations.

\subsection{Reconstructing Light Curves for Short Period Cepheids}

     The relatively large ranges of the Fourier amplitudes for Cepheids with periods less than 10 days (or $\log(P)<1.0$) is one reason for using the Fourier techniques to reconstruct the Cepheid light curves. From Figure \ref{fig:fig3} to \ref{fig:fig5}, it can be seen that the distribution of the Fourier amplitudes at periods longer than 10 days (or $\log(P) > 1.0$) show certain trends, which can be used to construct templates light curves as a function of period. However, the Fourier amplitudes for Cepheids with period less than 10 days do not show any obvious trends but scatter around certain ranges (as given in Table \ref{tab1}). For example, at a given short period, $A_1(V)$ may occupy the range from $\sim 0.1$ to $\sim 0.4$. Therefore, extra care has to be taken when applying template fitting to the Cepheids with periods less than 10 days. In fact, at periods shorter than $\log(P) < 0.85$ the template techniques do not apply \citep{ste96} but the Fourier techniques still holds good. 

     Although the extra-galactic Cepheids discovered in the past HST observations generally have period longer than 10 days, some short period period Cepheids have been discovered in Local Group galaxies IC 1613 \citep{dol01} \& Leo A \citep{dol02} with HST and ground based observations, respectively. Hence for observing Cepheids at short period, the Fourier techniques are still useful to reconstruct the light curves. With the new installation of ACS (Advanced Camera for Survey) in HST, the discovery of more short period Cepheids in other galaxies is probable.

\subsection{The Effect of Metallicity}

   One motivation for studying direct Fourier techniques to reconstruct Cepheid light curves is to deal with the possible metallicity dependence in Cepheids. \citet{ant00} shown that the metallicity may affect the shapes of light curves close to 10 days, based on the comparison of Cepheid light curves in two galaxies. The appropriate ranges of the Fourier parameters (as given in Table \ref{tab1}) are assumed to cover the possible ranges due to the different metallicity environments. In addition, the Fourier interrelations are shown to be only weakly dependent on the metallicity though this does not mean that the actual light curve shape is independent
of metallicity. 

    The template light curves defined by \citet{ste96} are obtained from the Galaxy ($Z=0.020$), LMC ($Z=0.008$) and SMC ($Z=0.004$), with sample of $\sim 30-45$ Cepheids in each galaxies. \citet{ste96} looked for and failed to find differences between the Galactic, LMC and SMC Cepheids in terms of their amplitude and light curve shape sufficient to change the estimation of mean magnitudes. However, much more data is available now. \citet{pac00} found statistically significant different amplitudes between the Galactic/LMC/SMC Cepheids in the period range $1.1 < \log (P) < 1.4$, in the sense that higher metallicity Cepheids have higher amplitudes. Moreover, in the period range $0.0 < \log (P) < 0.95$, the amplitudes ratio of $R_{21}$ and $R_{31}$ for Cepheids in LMC \citep{and87} and SMC \citep{and88} are found to be larger than the Galactic Cepheids. This may due to the different metallicity environments, at least for the case of the SMC \citep{buc94}. 
Further some established workers in the field have found
varying PC relations between the Galaxy, LMC and SMC and subsequently different PL
relations \citep{tam01,tam02}. Thus while it may be that template methods
yield a sufficiently accurate mean in varying metallicity environments, we contend that the issue of light curve shape and metallicity
needs to be revisited in the light of the data now currently available. 

\subsection{Applications in Determining the Cepheid Distances}

     An immediate application of these Fourier techniques is to reconstruct the Cepheid light curves in NGC 4258 and derive its distance modulus. NGC 4258 is a spiral galaxy that has an accurate geometrical distance of $7.2\pm0.3Mpc$ (corresponding to distance modulus of $29.28\pm0.09$) from water maser measurement \citep{her99}. By using the published data of 15 Cepheids discovered from HST observations \citep{new01} and following the H0KP procedures (including using the same PL relations and reddening correction), we derived a distance modulus of $29.38\pm0.06$ (random errors only) to NGC 4258, which is about $1.1\sigma$ away from water maser distance. If we apply the $-0.07mag$ correction due to the WFPC2 calibration of ``long vs short'' exposures (this is caused by the charge-transfer efficiency of WFPC2, see H0KP and reference therein for details), then the final distance modulus becomes $29.31\pm0.06$, which is consistent with the water maser distance. In addition, \citet{kan02} have also derived the Cepheid distance to NGC 4258 with the same techniques presented here but slightly different PL relations, the result of $29.36\pm0.06$ (random errors only) is also consistent to the water maser distance and the distance derived in this paper. 

When we apply these techniques to determine Cepheid distances to 16 galaxies using published photometry by the H0KP,
our distance moduli,
on average, are very similar to H0KP results, with a difference of $\sim0.01mag$ \citep{kan02}. On the other hands, application of these techniques to 3 Sandage-Tammann-Saha galaxies (NGC 3627: \citet{sah99}; NGC 3982: \citet{sah01a}; NGC 4527: \citet{sah01b}) yields shorter distance moduli of $0.07mags$ on average \citep{kan02}. The good agreements between the results in \citet{kan02} and in both H0KP and Sandage-Tammann-Saha galaxies suggest that our method is not too far of base.  

\acknowledgments

     This work has been supported by HST Grant AR08752.01-A. Part of SN work was performed under the auspices of the U.S. Department of Energy, National Nuclear Security Administration by the University of California, Lawrence Livermore National Laboratory under contract W7405-Eng-48. We thank an anonymous referee for many helpful suggestions for making the paper more
relevant.

\appendix
\section{The Fourier Intrarelations}

     The technique of Fourier interrelations, as described in Section 4, are the linear relations between the Fourier parameters in equation (1) in different wave bands. Similarly, there exist another group of linear relations between the Fourier parameters in the same wave band, known as the Fourier intrarelations. The reason for developing the Fourier intrarelations is same as in Fourier interrelation, mainly to reconstruct the I band light curves that only contain 4 epochs. Therefore, the Fourier intrarelations are the linear relation between the $1^{st}$ order Fourier parameter and the higher order Fourier parameters in one particular wave band, since 4 data points only permit the $1^{st}$ order Fourier expansion. The Fourier intrarelations have the following expression for either V or I bands:

     \begin{eqnarray}
       A_j & = & a_j + b_j A_1, \ \ j=2,\ldots 4 \\       
       \phi_j & = & c_j + d_j \phi_1, \ \ j=2,\ldots 4 
     \end{eqnarray}
 
     As in the case of Fourier interrelations, the coefficients are determined from the ``calibration set'' Cepheids. The results of the fits to the data with equation (A1) \& (A2) are presented in Table \ref{tab7}, which only listed out the $3^{rd}$ and $4^{th}$ order Fourier intrarelations. The $2^{nd}$ order Fourier intrarelations were omitted because for sparse, 12 epoch V band data, we can fit the data with $2^{nd}$ order Fourier expansion and expand to $4^{th}$ order with Fourier intrarelations to reconstruct V band light curves. Then we can use the Fourier interrelations to reconstruct the I band light curves from the V band light curves. 

     The plots for the Fourier intrarelations in ``calibrating set'' Cepheids are presented in Figure \ref{fig:fig21} \& \ref{fig:fig22} for Fourier amplitudes and phases, respectively. From these figures, though a relation clearly exists, it may not be linear though we show the best fit linear relation. In the plots of $A_3$ against $A_1$ for both V and I band, there are some stars which lie well below the best fit linear relation. The $A_4$ verses $A_1$ plots also show some evidence of a non-linearity. Since the error bars have been plotted on these diagrams, the trends described here are real. This non-linearity may due to differences in long and short period Cepheids.

    As in the case of Fourier interrelations, the Fourier intrarelations in OGLE LMC Cepheids have also been found, and presented in Table \ref{tab8}. The corresponding plots of the Fourier amplitudes and phases are given in Figure \ref{fig:fig23} \& \ref{fig:fig24}, respectively. The possible non-linearity of the Fourier intrarelations that seem on the ``calibrating set'' Cepheids also show up in these figures, as some stars which lie below the best fit linear regression ($A_3$ vs. $A_1$) and the indications of non-linearity in the $A_4$ vs. $A_1$ plots. Furthermore, by comparing the Fourier intrarelations in the ``calibrating set'' and OGLE LMC Cepheids, we see clearly that the slope of the best fit line increases from the ``calibrating set'' Cepheids to the LMC Cepheids. The difference in slope between ``calibrating set'' and LMC Cepheids found in the intrarelations plots is real and is probably attributable to the metallicity differences. 

    Due to the possible of non-linearity and metallicity dependence on the parent galaxy, the Fourier intrarelations are less preferable than the Fourier interrelations (Section 4) for reconstructing the I band light curves with only few data points available.

\clearpage

     \begin{table}
       \begin{center}
       \caption{The adopted ranges for the Fourier amplitudes, determined from Figure 3-5 for both V and I bands. \label{tab1}}
       \begin{tabular}{ccccc} \\ \tableline \tableline
         Period Ranges        & $A_1(V)$ & $A_2(V)$ & $A_3(V)$ & $A_4(V)$ \\ \tableline
         $0.0<\log(P)\leq1.0$ & 0.0-0.40 & 0.0-0.18 & 0.0-0.11 & 0.0-0.08 \\
         $1.0<\log(P)\leq1.1$ & 0.0-0.44 & 0.0-0.14 & 0.0-0.08 & 0.0-0.07 \\
         $1.1<\log(P)\leq1.2$ & 0.0-0.48 & 0.0-0.16 & 0.0-0.10 & 0.0-0.08 \\
         $1.2<\log(P)\leq1.3$ & 0.0-0.52 & 0.0-0.19 & 0.0-0.13 & 0.0-0.09 \\
         $1.3<\log(P)\leq1.4$ & 0.0-0.52 & 0.0-0.23 & 0.0-0.16 & 0.0-0.10 \\
         $1.4<\log(P)\leq1.5$ & 0.0-0.52 & 0.0-0.26 & 0.0-0.16 & 0.0-0.10 \\
         $1.5<\log(P)\leq1.6$ & 0.0-0.49 & 0.0-0.25 & 0.0-0.14 & 0.0-0.10 \\
         $1.6<\log(P)\leq1.7$ & 0.0-0.48 & 0.0-0.23 & 0.0-0.13 & 0.0-0.09 \\
         $1.7<\log(P)\leq1.8$ & 0.0-0.43 & 0.0-0.20 & 0.0-0.12 & 0.0-0.08 \\
         $1.8<\log(P)\leq1.9$ & 0.0-0.40 & 0.0-0.17 & 0.0-0.11 & 0.0-0.07 \\
         $1.9<\log(P)\leq2.0$ & 0.0-0.35 & 0.0-0.13 & 0.0-0.10 & 0.0-0.06 \\
          \tableline \tableline
         Period Ranges        & $A_1(I)$ & $A_2(I)$ & $A_3(I)$ & $A_4(I)$ \\ \tableline
         $0.0<\log(P)\leq1.0$ & 0.0-0.25 & 0.0-0.12 & 0.0-0.08 & 0.0-0.045 \\
         $1.0<\log(P)\leq1.1$ & 0.0-0.28 & 0.0-0.10 & 0.0-0.07 & 0.0-0.04 \\
         $1.1<\log(P)\leq1.2$ & 0.0-0.31 & 0.0-0.11 & 0.0-0.08 & 0.0-0.05 \\
         $1.2<\log(P)\leq1.3$ & 0.0-0.33 & 0.0-0.12 & 0.0-0.09 & 0.0-0.06 \\
         $1.3<\log(P)\leq1.4$ & 0.0-0.34 & 0.0-0.13 & 0.0-0.10 & 0.0-0.065 \\
         $1.4<\log(P)\leq1.5$ & 0.0-0.34 & 0.0-0.15 & 0.0-0.12 & 0.0-0.07 \\
         $1.5<\log(P)\leq1.6$ & 0.0-0.34 & 0.0-0.16 & 0.0-0.11 & 0.0-0.07 \\
         $1.6<\log(P)\leq1.7$ & 0.0-0.31 & 0.0-0.15 & 0.0-0.10 & 0.0-0.065 \\
         $1.7<\log(P)\leq1.8$ & 0.0-0.27 & 0.0-0.13 & 0.0-0.09 & 0.0-0.06 \\
         $1.8<\log(P)\leq1.9$ & 0.0-0.24 & 0.0-0.11 & 0.0-0.08 & 0.0-0.055 \\
         $1.9<\log(P)\leq2.0$ & 0.0-0.21 & 0.0-0.10 & 0.0-0.07 & 0.0-0.05 \\
         \tableline
       \end{tabular}
       \end{center}
     \end{table}

\clearpage

    \begin{table}
      \begin{center}
      \caption{The coefficients of the Fourier Interrelations, as determined from ``calibrating set'', for fundamental mode Cepheids. Coefficients are corresponding to those defined in equation (3) \& (4). Parameter $\chi^2$ characterizes the goodness of fit: $\chi^2=\sum_i (y_i-\alpha-\beta x_i)^2/(\sigma_{y,i}^2+\beta^2 \sigma_{x,i}^2)$ \label{tab2}}
      \begin{tabular}{lccc} \\ \tableline \tableline
        Relation & $\alpha$            & $\beta$  & $\chi^2$  \\ \tableline
        $A_1$    & $-0.006\pm0.002$ & $0.643\pm 0.007$ & $2.69\times 10^2$ \\
        $A_2$    & $ 0.001\pm0.001$ & $0.600\pm 0.011$ & $9.74\times 10^2$ \\
        $A_3$    & $ 0.000\pm0.001$ & $0.645\pm 0.023$ & $6.05\times 10^1$ \\
        $A_4$    & $ 0.000\pm0.001$ & $0.631\pm 0.039$ & $4.85\times 10^1$ \\  \tableline \tableline
        Relation & $\gamma$            & $\eta$   & $\chi^2$  \\ \tableline
        $\phi_1$ & $-0.178\pm 0.033$ & $0.996\pm 0.001$ & $2.21\times 10^2$  \\
        $\phi_2$ & $-0.048\pm 0.010$ & $1.005\pm 0.003$ & $6.89\times 10^1$  \\
        $\phi_3$ & $-0.015\pm 0.021$ & $1.003\pm 0.004$ & $1.02\times 10^2$  \\
        $\phi_4$ & $-0.001\pm 0.034$ & $1.004\pm 0.006$ & $7.10\times 10^1$ \\
        \tableline
      \end{tabular}
      \end{center}
    \end{table}

\clearpage

    \begin{table}
      \begin{center}
      \caption{The coefficients of the Fourier Interrelations for OGLE LMC (fundamental mode) Cepheids. The columns are the same as in Table \ref{tab2}. \label{tab3}}
      \begin{tabular}{lccc} \\ \tableline \tableline
        Relation & $\alpha$            & $\beta$  & $\chi^2$  \\ \tableline
        $A_1$    & $ 0.004\pm0.005$ & $0.601\pm0.002$ & $2.64\times 10^3$ \\
        $A_2$    & $ 0.001\pm0.004$ & $0.601\pm0.004$ & $1.68\times 10^3$ \\
        $A_3$    & $ 0.000\pm0.001$ & $0.606\pm0.006$ & $1.36\times 10^3$ \\
        $A_4$    & $-0.001\pm0.001$ & $0.625\pm0.015$ & $1.16\times 10^3$ \\  \tableline \tableline
        Relation & $\gamma$            & $\eta$   & $\chi^2$  \\ \tableline
        $\phi_1$ & $-0.208\pm0.002$ & $1.001\pm0.001$ & $4.79\times 10^3$  \\
        $\phi_2$ & $-0.067\pm0.004$ & $1.001\pm0.001$ & $1.49\times 10^3$  \\
        $\phi_3$ & $-0.029\pm0.009$ & $1.001\pm0.001$ & $1.20\times 10^3$  \\
        $\phi_4$ & $-0.063\pm0.022$ & $1.008\pm0.003$ & $9.15\times 10^2$ \\
        \tableline
      \end{tabular}
      \end{center}
    \end{table}
    
\clearpage

    \begin{table}
      \begin{center}
      \caption{The coefficients of the Fourier Interrelations for OGLE SMC (fundamental mode) Cepheids. The columns are the same as in Table \ref{tab2}. \label{tab4}}
      \begin{tabular}{lccc} \\ \tableline \tableline
        Relation & $\alpha$            & $\beta$  & $\chi^2$  \\ \tableline
        $A_1$    & $ 0.003\pm0.001$ & $0.608\pm0.002$ & $3.87\times 10^3$ \\
        $A_2$    & $-0.000\pm0.001$ & $0.623\pm0.003$ & $2.91\times 10^3$ \\
        $A_3$    & $-0.001\pm0.001$ & $0.650\pm0.005$ & $2.65\times 10^3$ \\
        $A_4$    & $-0.002\pm0.001$ & $0.647\pm0.010$ & $2.28\times 10^3$ \\  \tableline \tableline
        Relation & $\gamma$            & $\eta$   & $\chi^2$  \\ \tableline
        $\phi_1$ & $-0.214\pm0.003$ & $1.000\pm0.001$ & $5.46\times 10^3$  \\
        $\phi_2$ & $-0.075\pm0.004$ & $1.001\pm0.001$ & $3.06\times 10^3$  \\
        $\phi_3$ & $-0.018\pm0.008$ & $1.000\pm0.001$ & $2.08\times 10^3$  \\
        $\phi_4$ & $-0.089\pm0.016$ & $1.008\pm0.002$ & $1.54\times 10^3$ \\
        \tableline
      \end{tabular}
      \end{center}
    \end{table}
 
\clearpage

    \begin{table}
      \begin{center}
      \caption{The results of error analysis from simulations in V band. The Gaussian parameters $\mu$ and $\sigma$ are mean off-sets and errors, respectively, for the means and the Fourier amplitudes. \label{tab5}}
      \begin{tabular}{lccccc} \\ \tableline \tableline
        Simulation & $\mu_{mean}$ & $\mu_{A1}$ & $\mu_{A2}$ & $\mu_{A3}$ & $\mu_{A4}$  \\ \tableline
        1          & 0.0014       & -0.0121    & -0.0063    & 0.0002     & 0.0063 \\
        2          & 0.0006       & -0.0174    & -0.0062    & 0.0013     & 0.0095 \\
        3          & -0.0018      & -0.0226    & -0.0154    & -0.0071    & 0.0059 \\  \tableline \tableline
	Simulation & $\sigma_{mean}$ & $\sigma_{A1}$ & $ \sigma_{A2}$ & $\sigma_{A3}$ & $\sigma_{A4}$  \\ \tableline
        1          & 0.0312          & 0.0355        & 0.0330         & 0.0259        & 0.0210 \\
        2          & 0.0337          & 0.0393        & 0.0332         & 0.0280        & 0.0228 \\
        3          & 0.0256          & 0.0324        & 0.0272         & 0.0242        & 0.0172 \\  \tableline \tableline
        \tableline
      \end{tabular}
      \end{center}
    \end{table}

\clearpage

    \begin{table}
      \begin{center}
      \caption{The results of error analysis from simulations in I band. The Gaussian parameters $\mu$ and $\sigma$ are mean off-sets and errors, respectively, for the means and the Fourier amplitudes. \label{tab6}}
      \begin{tabular}{lccccc} \\ \tableline \tableline
        Simulation & $\mu_{mean}$ & $\mu_{A1}$ & $\mu_{A2}$ & $\mu_{A3}$ & $\mu_{A4}$  \\ \tableline
        1          & 0.0012       & -0.0071    & -0.0053    & 0.0018     & 0.0048  \\
        2          & 0.0010       & -0.0109    & -0.0052    & 0.0020     & 0.0064  \\
        3          & -0.0042      & -0.0132    & -0.0039    & 0.0018     & -0.0004 \\  \tableline \tableline
	Simulation & $\sigma_{mean}$ & $\sigma_{A1}$ & $ \sigma_{A2}$ & $\sigma_{A3}$ & $\sigma_{A4}$  \\ \tableline
        1          & 0.0283          & 0.0226        & 0.0195         & 0.0158        & 0.0130 \\
        2          & 0.0348          & 0.0253        & 0.0206         & 0.0175        & 0.0145 \\
        3          & 0.0251          & 0.0208        & 0.0163         & 0.0156        & 0.0108 \\  \tableline \tableline
        \tableline
      \end{tabular}
      \end{center}
    \end{table}

\clearpage

     \begin{table}
      \begin{center}
      \caption{The coefficients of the Fourier Intrarelations, as determined from ``calibrating set'', for fundamental mode Cepheids. Coefficients are corresponding to those defined in equation (A1) \& (A2). $\chi^2$ parameter is same as in Table \ref{tab2}. \label{tab7}}
      \begin{tabular}{lccc} \\ \tableline \tableline
        Relation & $a$            & $b$  & $\chi^2$  \\ \tableline
        $A_3(V)$    & $-0.043\pm0.002$ & $0.281\pm0.006$ & $3.97\times 10^2$ \\
        $A_4(V)$    & $-0.021\pm0.002$ & $0.139\pm0.006$ & $2.97\times 10^2$ \\
        $A_3(I)$    & $-0.025\pm0.002$ & $0.275\pm0.010$ & $2.77\times 10^2$ \\
        $A_4(I)$    & $-0.014\pm0.002$ & $0.145\pm0.010$ & $1.98\times 10^2$ \\  \tableline \tableline
        Relation & $c$            & $d$   & $\chi^2$  \\ \tableline
        $\phi_3(V)$ & $2.544\pm0.020$ & $3.017\pm0.008$ & $2.42\times 10^3$  \\
        $\phi_4(V)$ & $0.202\pm0.034$ & $4.061\pm0.013$ & $1.40\times 10^3$  \\
        $\phi_3(I)$ & $3.122\pm0.027$ & $3.043\pm0.011$ & $1.72\times 10^3$  \\
        $\phi_4(I)$ & $0.887\pm0.044$ & $4.122\pm0.020$ & $9.02\times 10^3$ \\
        \tableline
      \end{tabular}
      \end{center}
    \end{table}

\clearpage

    \begin{table}
      \begin{center}
      \caption{The coefficients of the Fourier Intrarelations for OGLE LMC (fundamental mode) Cepheids. The columns are same as in Table \ref{tab4}. \label{tab8}}
      \begin{tabular}{lccc} \\ \tableline \tableline
        Relation & $a$            & $b$  & $\chi^2$  \\ \tableline
        $A_3(V)$    & $-0.051\pm0.001$ & $0.364\pm0.005$ & $2.53\times 10^3$ \\
        $A_4(V)$    & $-0.034\pm0.002$ & $0.214\pm0.005$ & $1.71\times 10^3$ \\
        $A_3(I)$    & $-0.031\pm0.001$ & $0.355\pm0.002$ & $1.42\times 10^4$ \\
        $A_4(I)$    & $-0.019\pm0.001$ & $0.198\pm0.002$ & $8.24\times 10^3$ \\  \tableline \tableline
        Relation & $c$            & $d$   & $\chi^2$  \\ \tableline
        $\phi_3(V)$ & $2.076\pm0.029$ & $3.106\pm0.009$ & $2.59\times 10^3$  \\
        $\phi_4(V)$ & $0.072\pm0.055$ & $4.084\pm0.017$ & $1.74\times 10^3$  \\
        $\phi_3(I)$ & $2.901\pm0.011$ & $3.042\pm0.003$ & $1.57\times 10^4$  \\
        $\phi_4(I)$ & $1.193\pm0.019$ & $3.995\pm0.006$ & $7.81\times 10^3$ \\
        \tableline
      \end{tabular}
      \end{center}
    \end{table}

\clearpage

\clearpage

\clearpage

    \begin{figure}
      \caption{The distributions of the Fourier amplitudes for OGLE LMC Cepheids in V band. Total number of the Cepheids is 758, which includes 648 Cepheids with well constructed light curves (crosses) and 110 Cepheids with poorly constructed light curves (triangles). The Cepheids in the ``calibrating set'' are shown as filled circles, for comparison.\label{fig:fig3}}
    \end{figure}

\clearpage

     \begin{figure}
       \caption{The distributions of the Fourier amplitudes for OGLE LMC Cepheids in V band (crosses) with well constructed light curves. The Cepheids in the ``calibrating set'' are shown as filled circles, for comparison.\label{fig:fig4}}
     \end{figure} 

\clearpage

     \begin{figure}
       \caption{The distributions of the Fourier amplitudes for OGLE LMC Cepheids in I band are shown as crosses. Some of the suspicious outliers are shown as triangles and labelled. The Cepheids in ``calibrating set'' are shown as filled circles, for comparison.\label{fig:fig5}}
     \end{figure} 

\clearpage

     \begin{figure}
       \plottwo{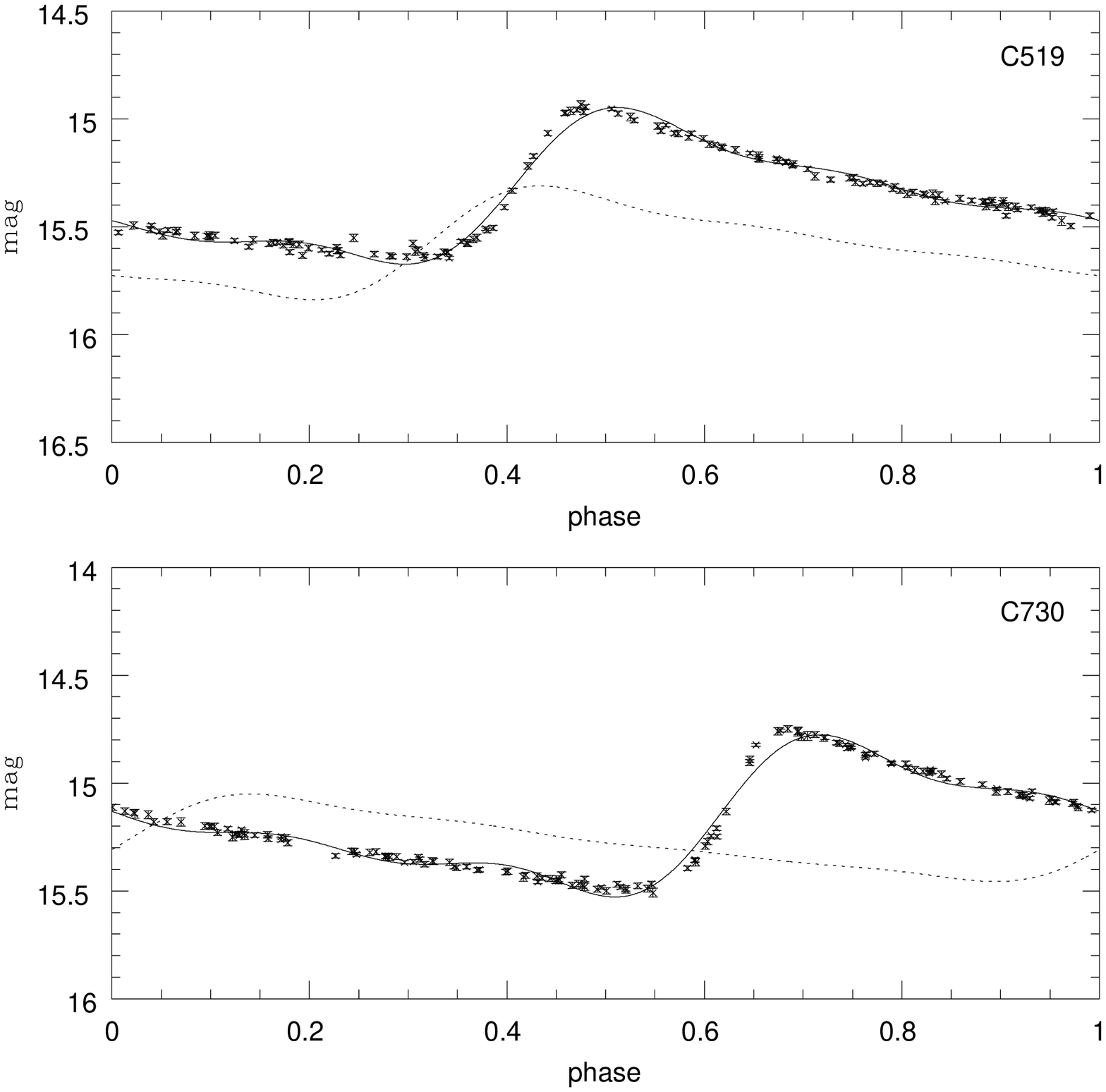}{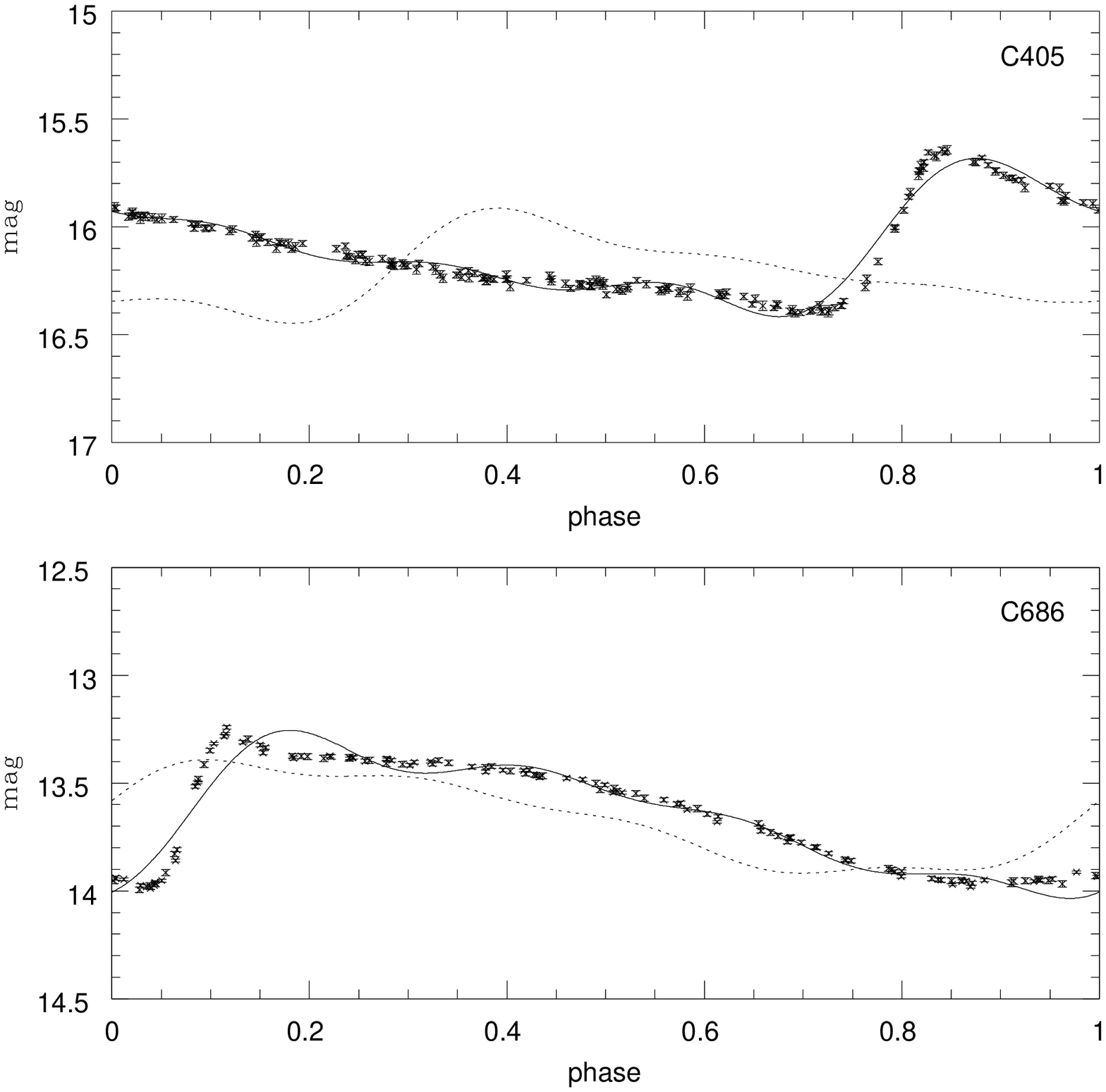}
       \caption{Examples of the constructed light curves for the outliers in Figure \ref{fig:fig5}. The dashed lines are the light curves for Cepheid with similar periods, for comparison  .\label{fig:fig6}}
     \end{figure} 

\clearpage

     \begin{figure}
       \plottwo{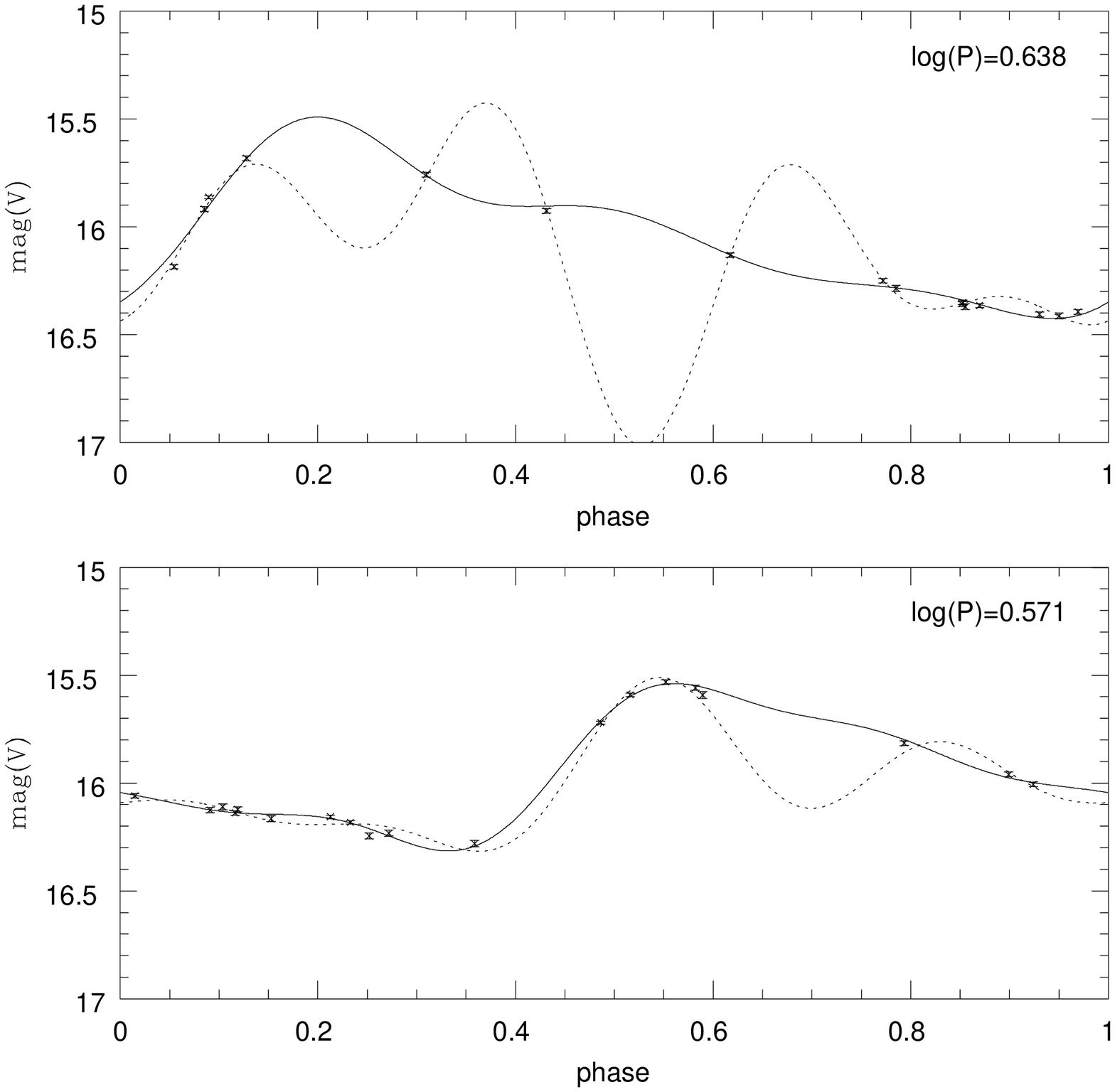}{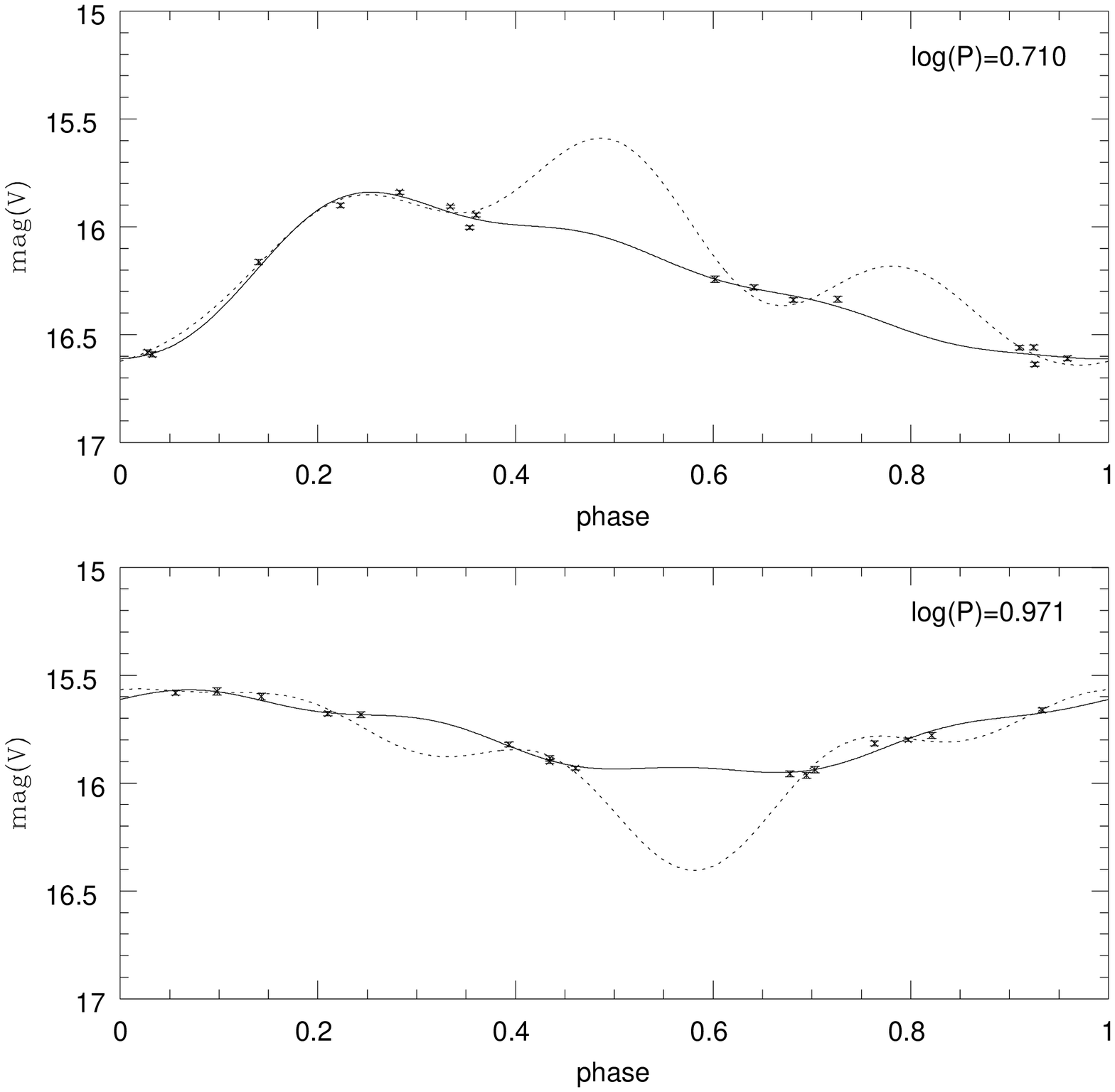}
        \caption{Examples of the reconstructed light curves fitted with the ranges of Fourier amplitudes given in Table \ref{tab1}. These light curves are indicated in solid curves. The dotted curves are the light curves fitted with initial ranges of Fourier amplitudes (i.e. $A_0:0-0.7;\ A_1,A_2,A_3:0-0.5$). The periods of the OGLE LMC Cepheids are also given in upper-right corner. Original data points are also indicated.\label{fig:fig7}}
     \end{figure} 

\clearpage

     \begin{figure}
      \plottwo{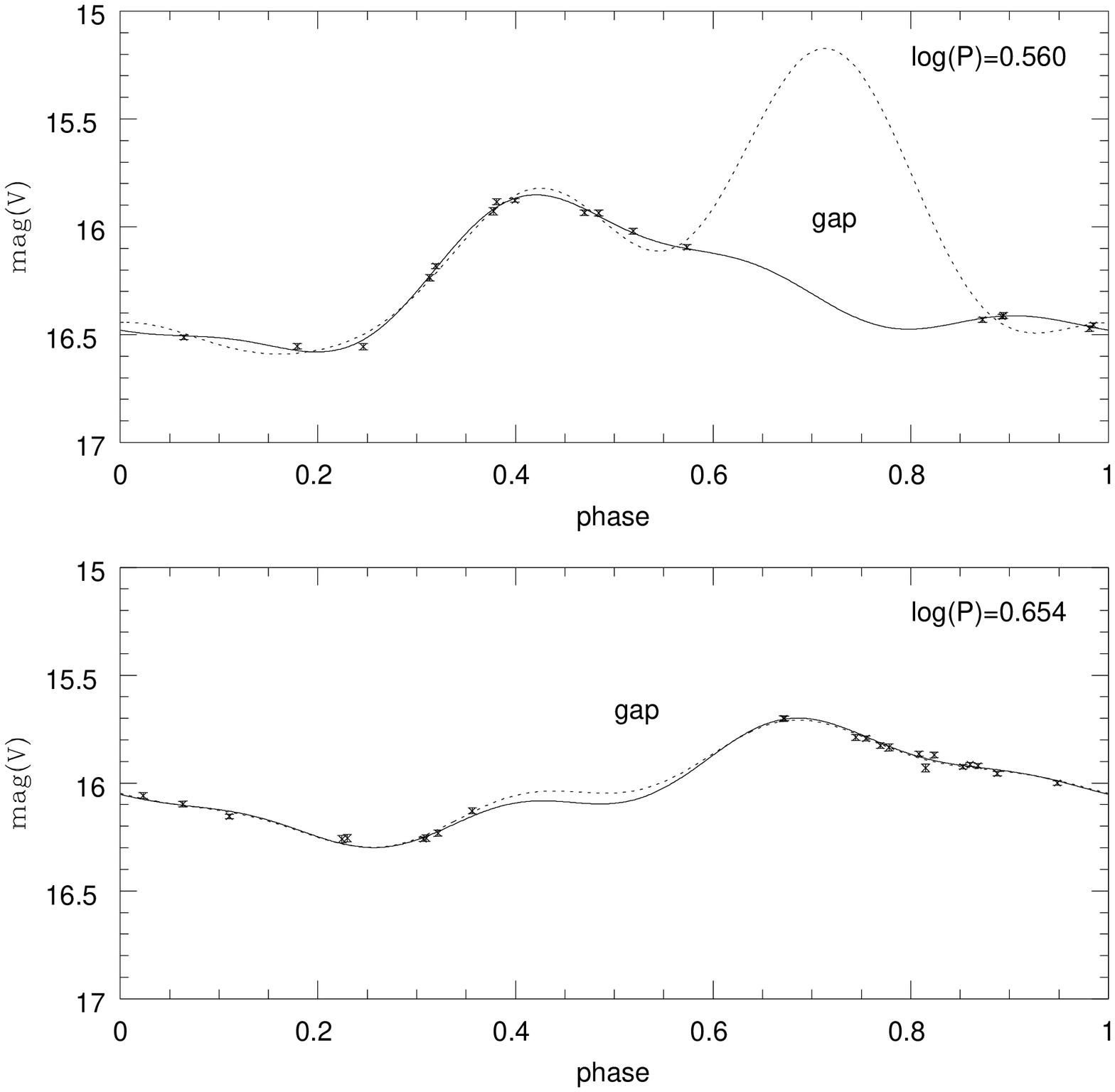}{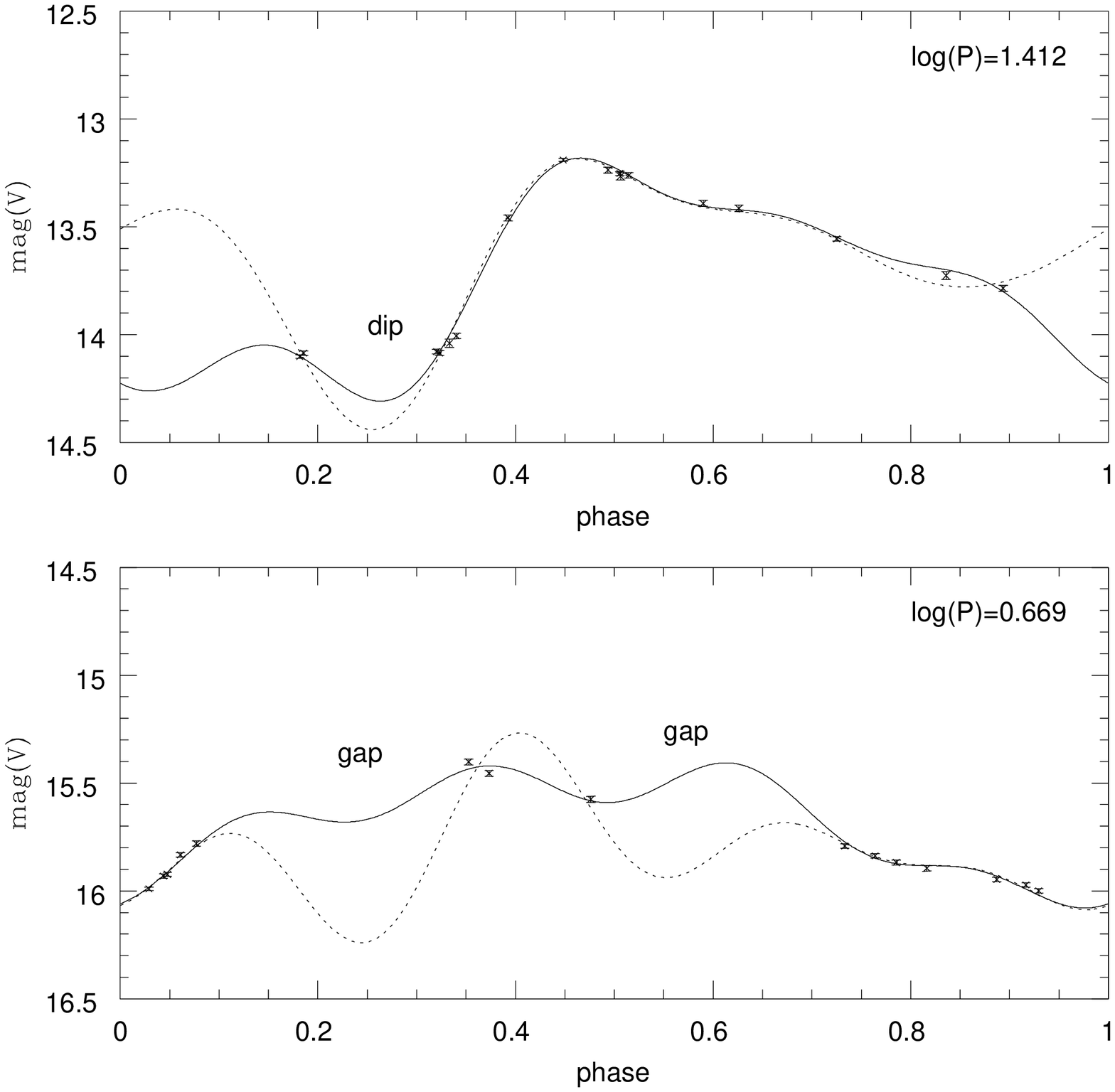}
       \caption{Examples of the reconstructed light curves that exhibit numerical bumps/dips, regardless of fitting the data with smaller (solid curves, as given in Table \ref{tab1}) or larger (dotted curves) ranges of the Fourier amplitudes. This is mainly due to the bad phase coverage in the data, with gaps in between the data points. Original data points are also indicated.\label{fig:fig8}}
     \end{figure} 

\clearpage

    \begin{figure}
      \caption{The Interrelations among the Fourier amplitudes in the ``calibrating set'' Cepheids. Crosses represent corresponding Fourier amplitudes and standard errors for each Cepheids. The dashed lines are the best-fit straight lines to the data, with the slopes and zero points given in Table \ref{tab2}. Solid and open squares are for bump and non-bump Cepheids.\label{fig:fig9}}
    \end{figure}

\clearpage

    \begin{figure}
      \caption{Same as Figure \ref{fig:fig9}, except for the Interrelations among the Fourier phases in the ``calibrating set'' Cepheids.\label{fig:fig10}}
    \end{figure}

\clearpage

    \begin{figure}
      \caption{The Interrelations among the Fourier amplitudes in the OGLE LMC (fundamental mode) Cepheids. The solid lines are the best-fit straight lines to the data, with the slopes and zero points given in Table \ref{tab3}. The dashed lines are the best-fit model for ``calibrating set'' Cepheids, for comparison.\label{fig:fig11}}
    \end{figure}

\clearpage

    \begin{figure}
      \caption{Same as Figure \ref{fig:fig11}, except for the Interrelations among the Fourier phases in the OGLE LMC Cepheids.\label{fig:fig12}}
    \end{figure}

\clearpage

   \begin{figure}
      \caption{The Interrelations among the Fourier amplitudes in the OGLE SMC (fundamental mode) Cepheids. The solid lines are the best-fit straight lines to the data. The dashed lines are the best-fit model for ``calibrating set'' Cepheids, for comparison.\label{fig:fig13}}
    \end{figure}

\clearpage

    \begin{figure}
      \caption{Same as Figure \ref{fig:fig13}, except for the Interrelations among the Fourier phases in the OGLE SMC Cepheids.\label{fig:fig14}}
    \end{figure}
  
\clearpage

     \begin{figure}
       \plottwo{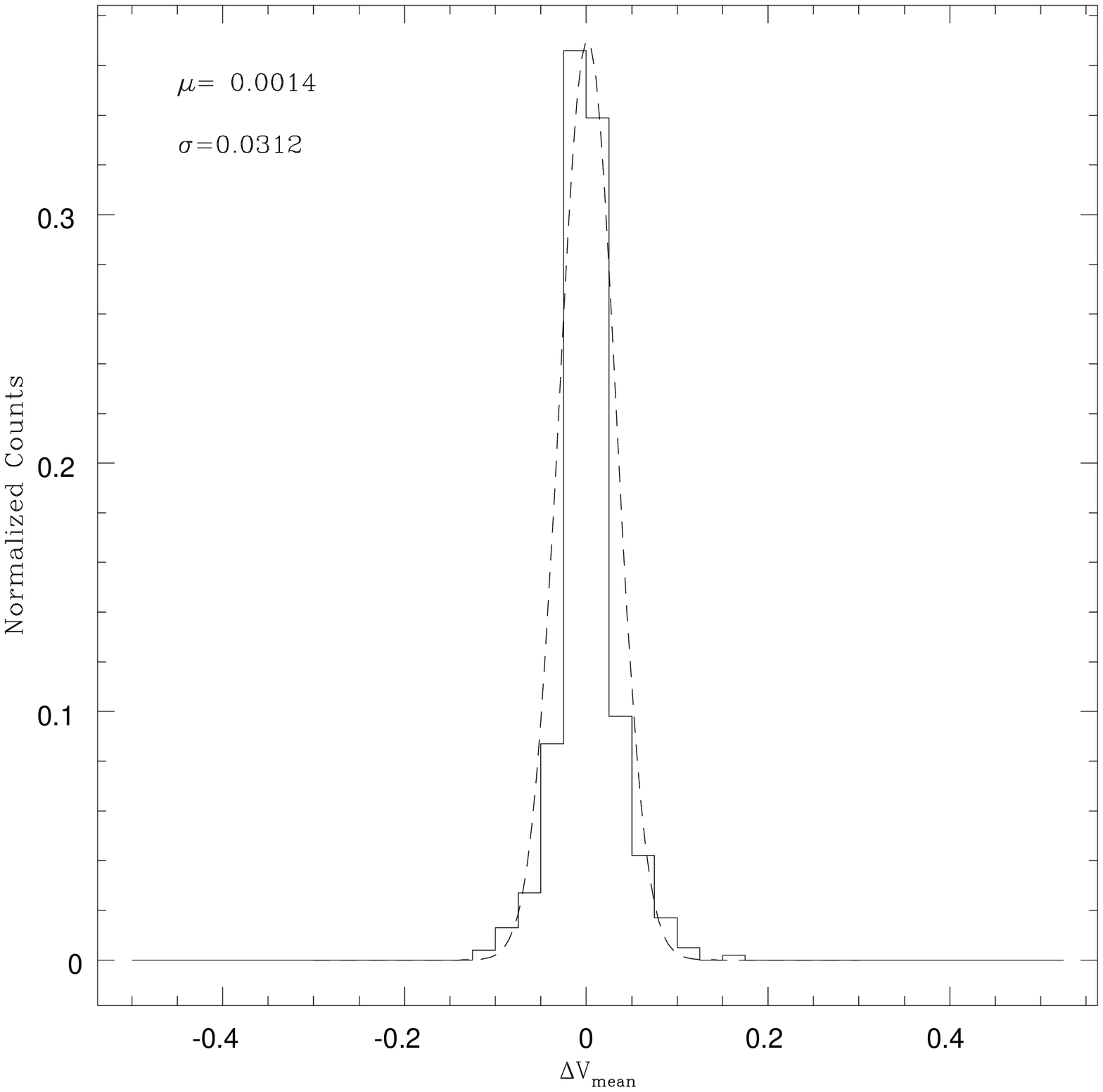}{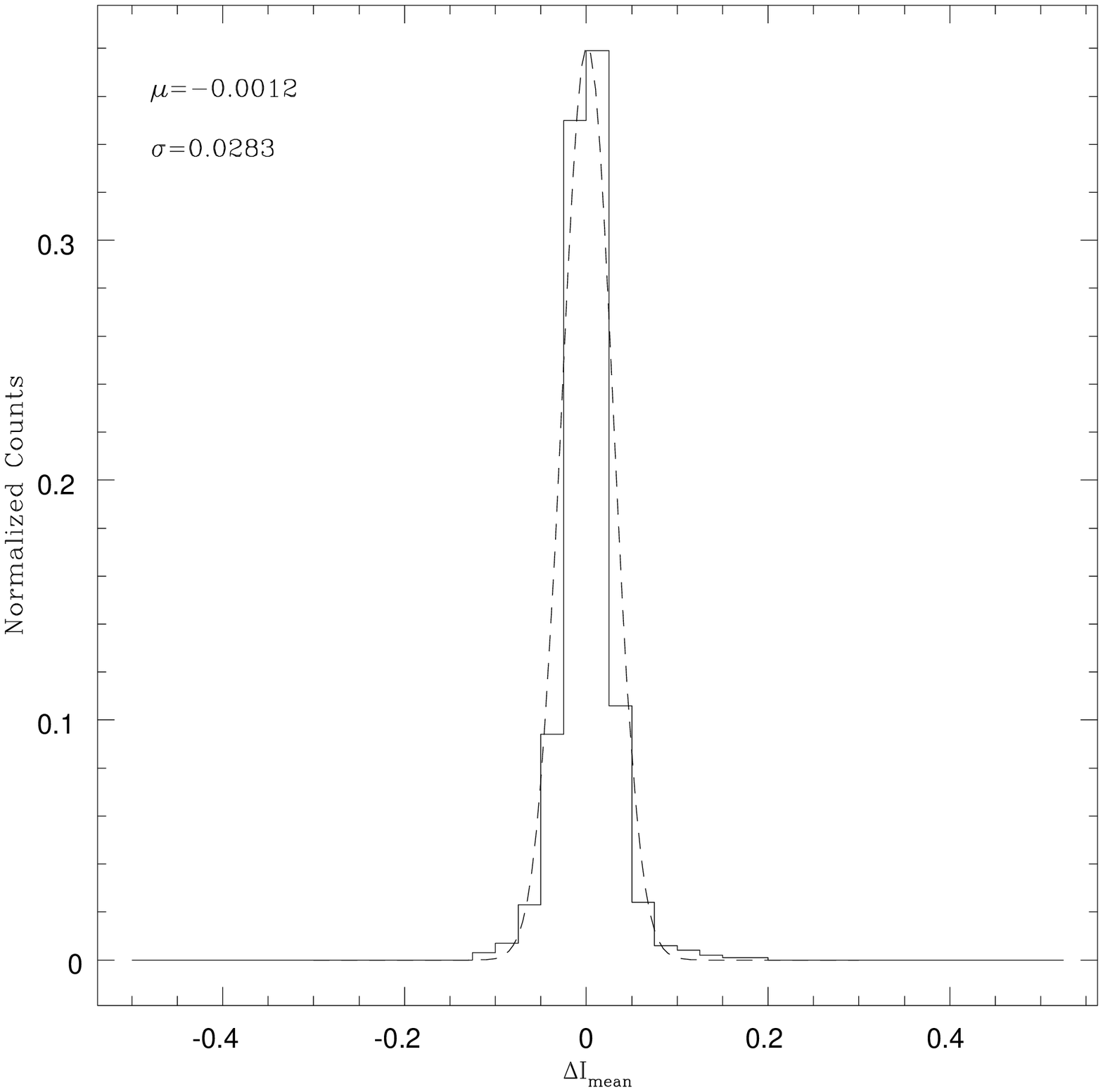}
       \caption{Histograms of offsets for mean magnitudes in OGLE LMC Cepheids from light curve reconstruction procedures. The x-axes are the offsets of the simulated means from the original values, and the y-axes are the normalized counts. The dashed curves are the fitted Gaussian distribution to the histograms, with the parameters given in upper-left corners. Left and right panels are for V and I bands, respectively. \label{fig:fig15}}
     \end{figure}
     
\clearpage

     \begin{figure}
       \plotone{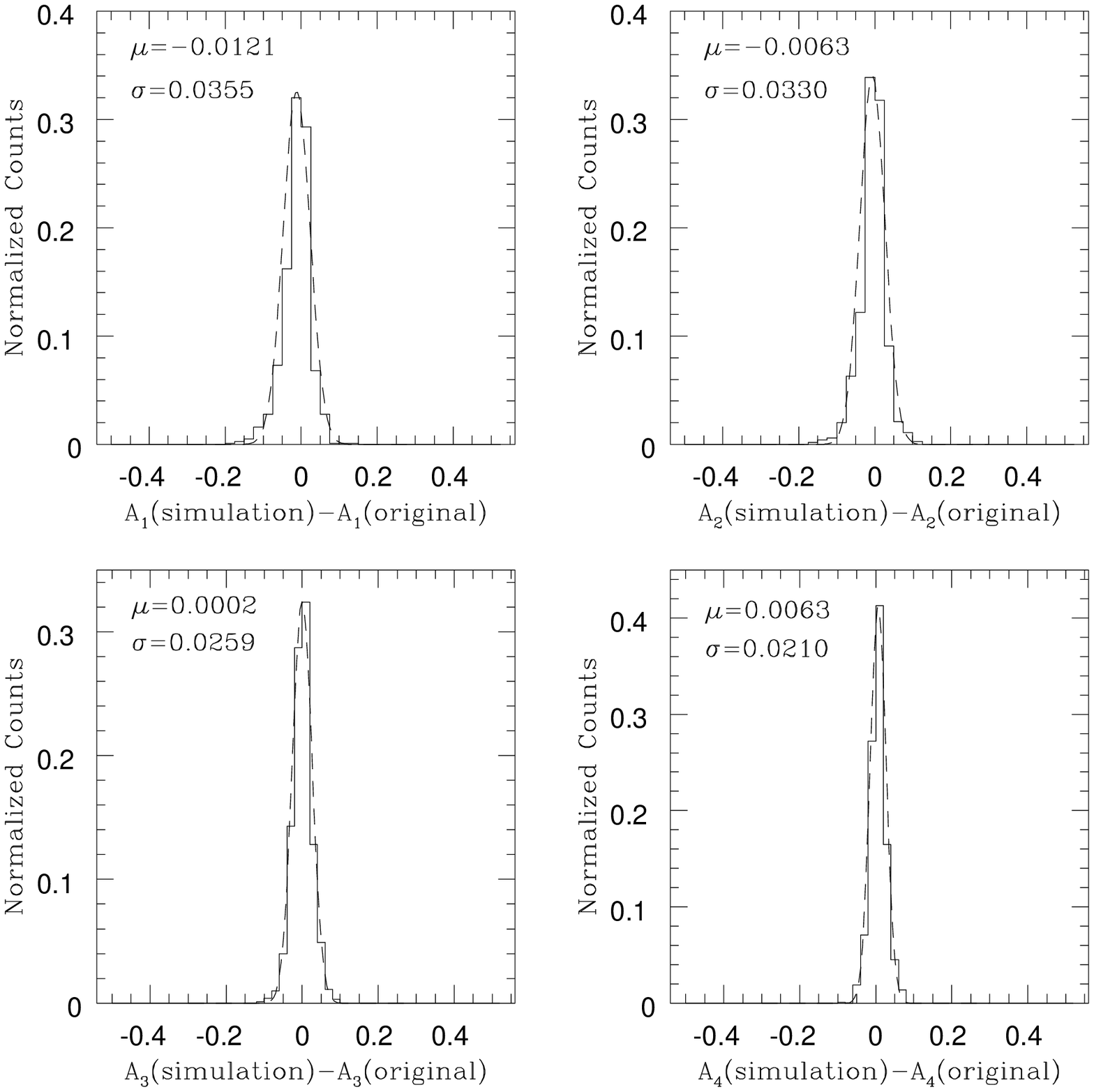}
       \caption{Histograms of offsets for OGLE LMC Cepheids in light curve reconstruction procedure in V band. The dashed curves are the fitted Gaussian distribution to the histograms, with the parameters given in upper-left corners. \label{fig:fig16}}
     \end{figure}

\clearpage

     \begin{figure}
       \plotone{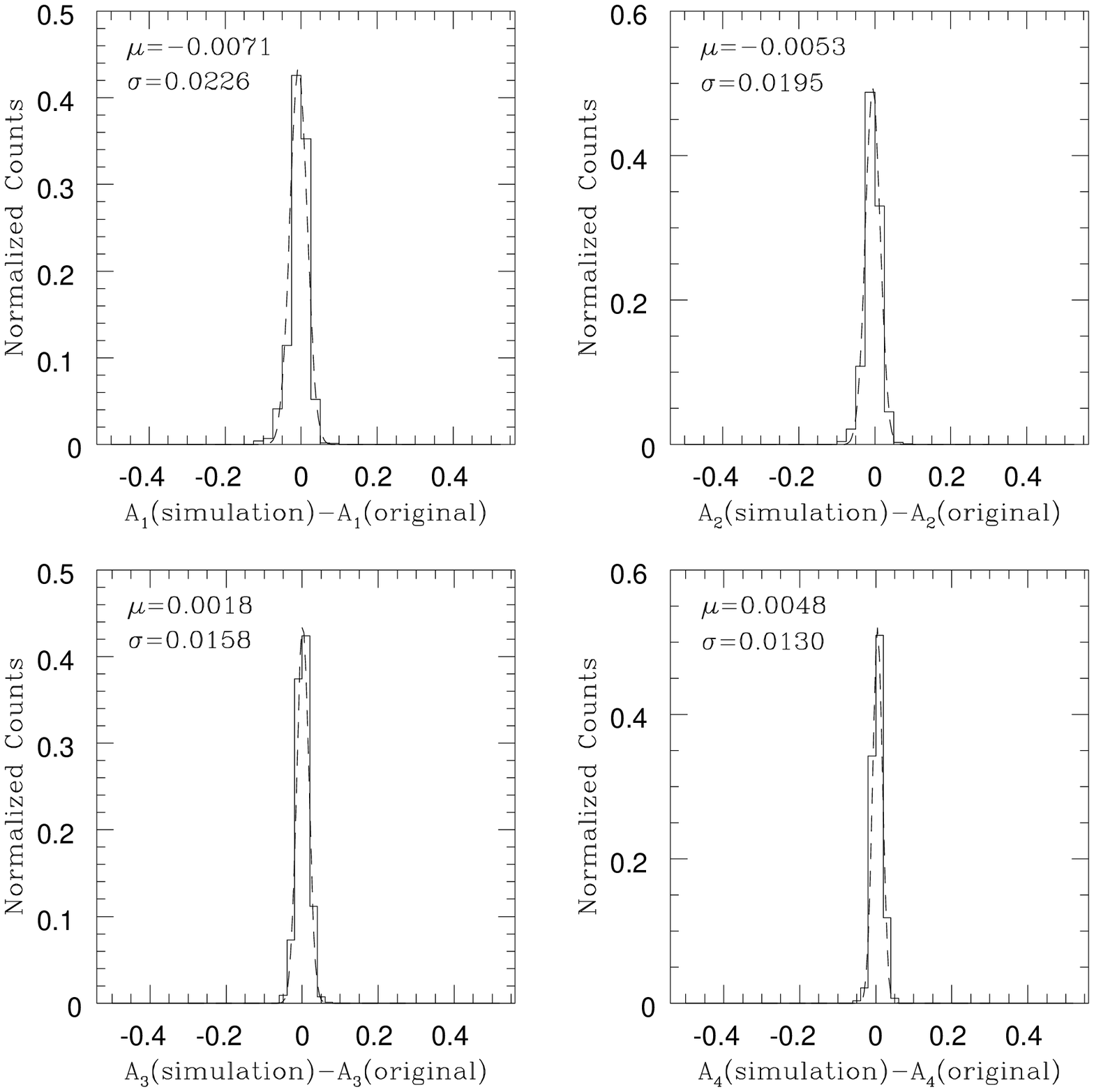}
       \caption{Histograms of offsets for OGLE LMC Cepheids in light curve reconstruction procedure in I band. The dashed curves are the fitted Gaussian distribution to the histograms, with the parameters given in upper-left corners. \label{fig:fig17}}
     \end{figure}

\clearpage

     \begin{figure}
       \caption{Plot of $R_{21}=A_2/A_1$ in V band for simulated data (crosses). The crosses with triangles are the simulated data points with either the values of $A_1$ or $A_2$ is  $\sim 2.0\sigma$ away from the original values. The filled circles are original OGLE LMC Cepheid data. \label{fig:fig18}}
     \end{figure}

\clearpage

     \begin{figure}
       \plottwo{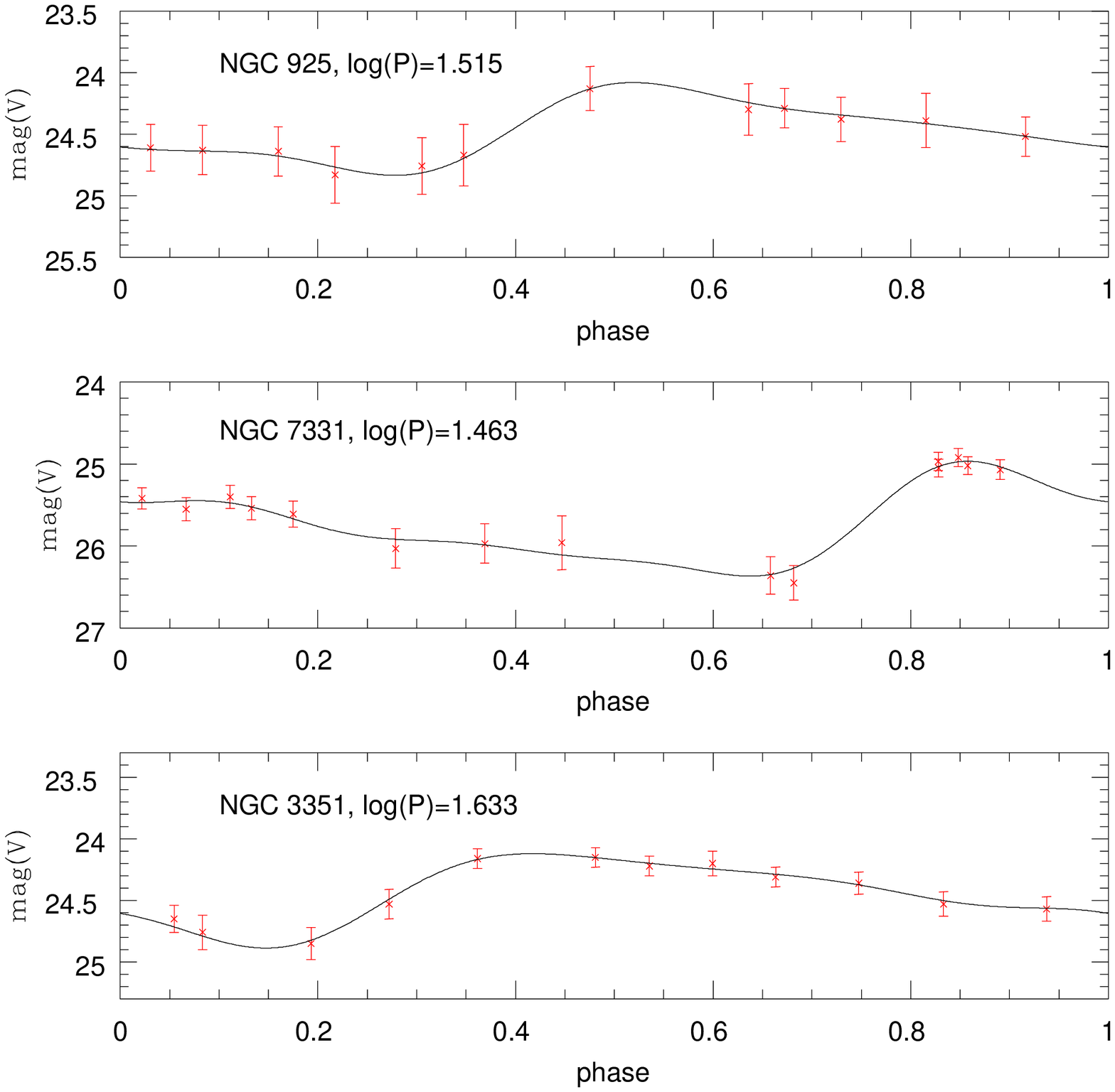}{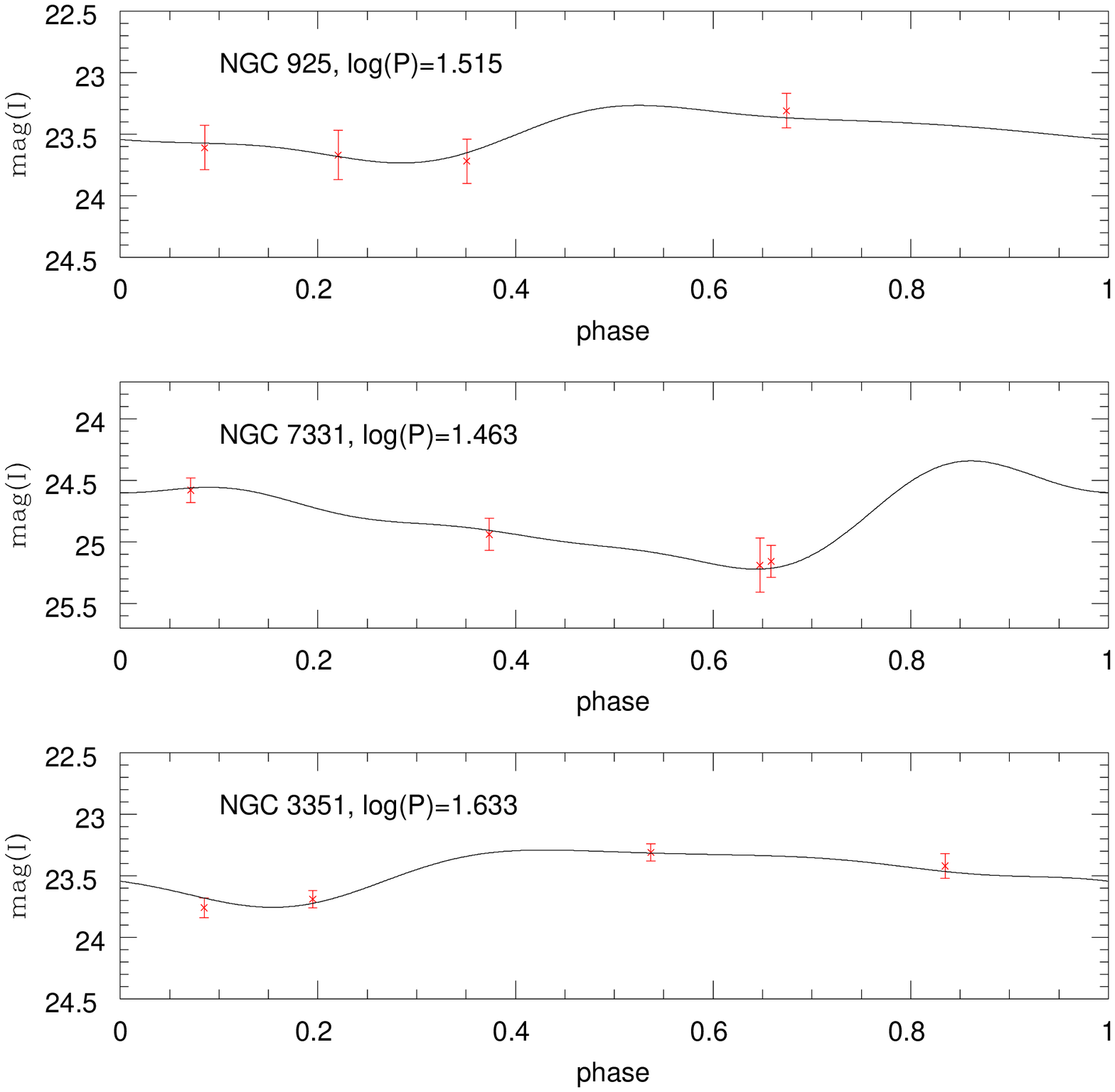}
       \caption{Examples of the reconstructed light curves in nearby galaxies. The V band light curves (left panel) are reconstructed via $4^{th}$ order Fourier expansion and the I band light curves (right panel) are reconstruct via the Fourier interrelations.\label{fig:fig19}}
     \end{figure}

\clearpage

     \begin{figure}
       \plottwo{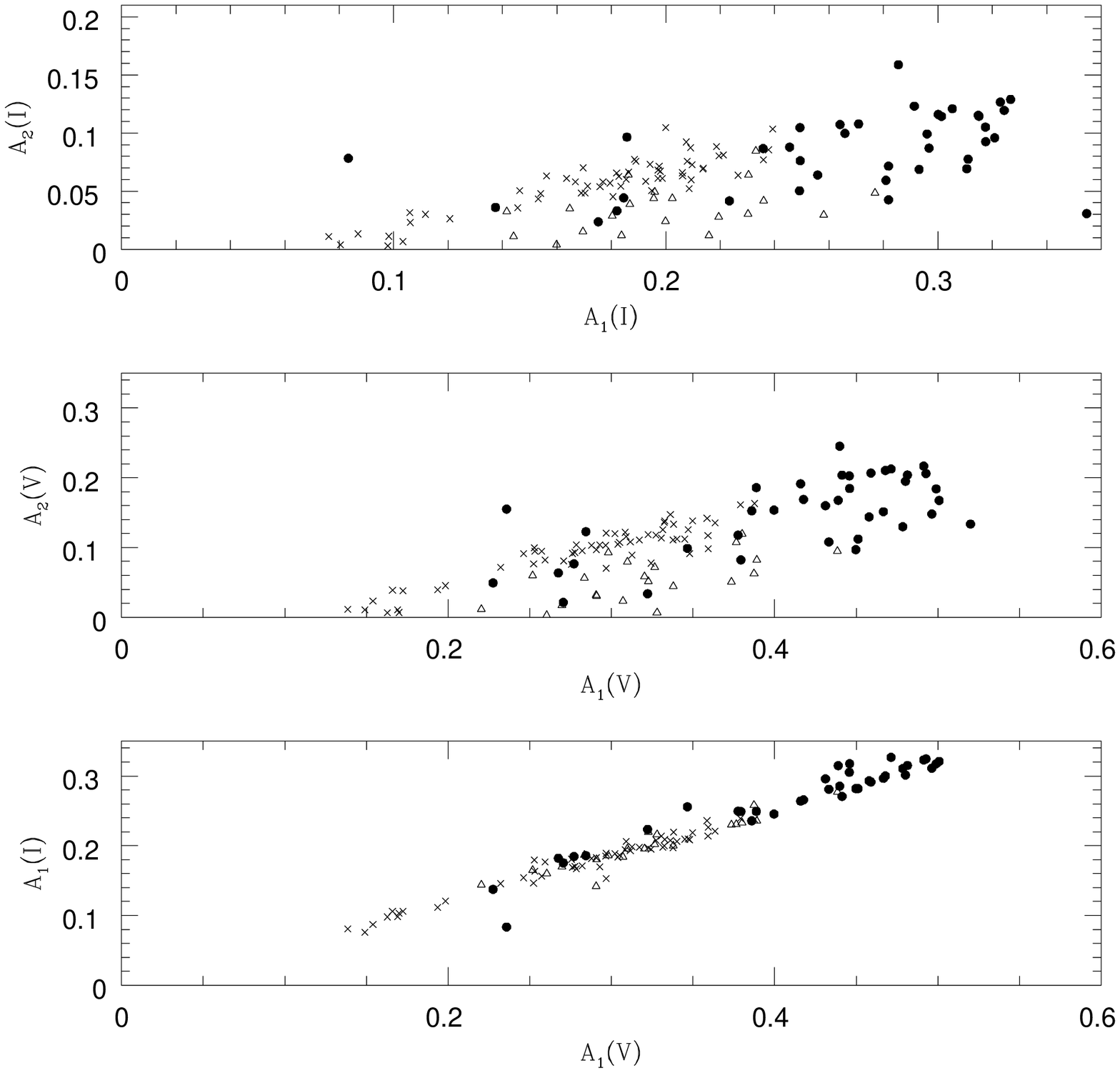}{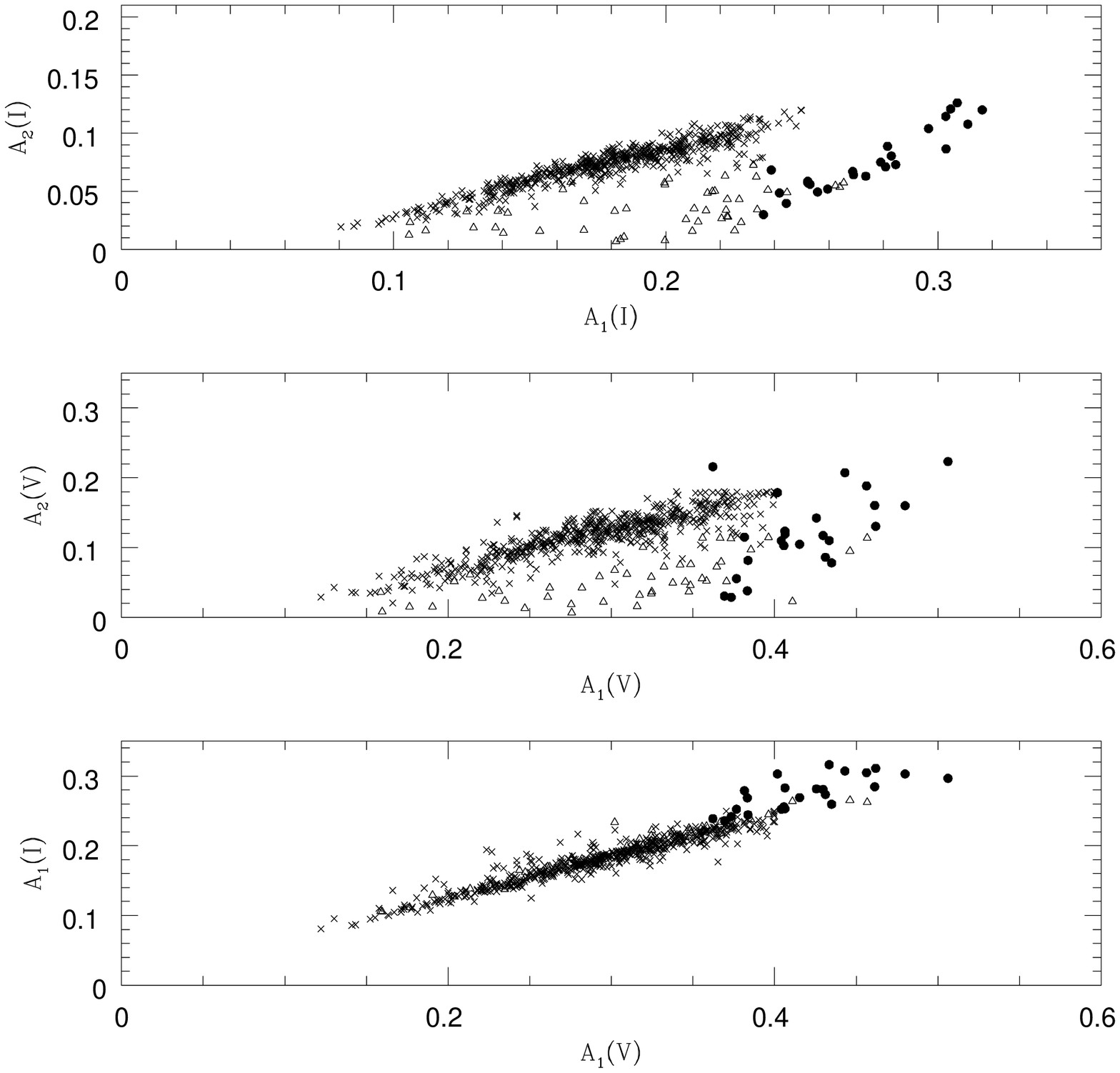}
       \caption{Comparisons of the correlations for $A_1-A_2$ and $A_1(V)-A_1(I)$. Crosses are for Cepheid with period shorter than 8 days; triangles are for Cepheids with period in between 8 to 14 days; and filled circles are for Cepheids with period longer than 14 days. Error bars are omitted for clarity. \textit{Left:} ``Calibrating set'' Cepheids; \textit{Right:} OGLE LMC Cepheids. \label{fig:fig20}}
     \end{figure}

\clearpage

    \begin{figure}
      \caption{The Intrarelations among the Fourier amplitudes in the ``calibrating set'' Cepheids. The dashed lines are the best-fit straight lines to the data.\label{fig:fig21}}
    \end{figure}

\clearpage

    \begin{figure}
      \caption{Same as Figure \ref{fig:fig21}, except for the Intrarelations among the Fourier phases in the ``calibrating set'' Cepheids.\label{fig:fig22}}
    \end{figure}

\clearpage

    \begin{figure}
      \caption{The Intrarelations among the Fourier amplitudes in the OGLE LMC (fundamental mode) Cepheids. The solid lines are the best-fit straight lines to the data. The dashed lines are the best-fit model for ``calibrating set'' Cepheids, for comparison.\label{fig:fig23}}
    \end{figure}

\clearpage

    \begin{figure}
      \caption{Same as Figure \ref{fig:fig23}, except for the Intrarelations among the Fourier phases in the OGLE LMC Cepheids.\label{fig:fig24}}
    \end{figure}


\begin{thebibliography}{}
\bibitem[Andreasen \& Petersen, 1987]{and87} Andreasen, G. \& Petersen, J., 1987, \aap, 180, 129
\bibitem[Andreasen, 1988]{and88} Andreasen, G., 1988, \aap, 191, 71
\bibitem[Antonello \& Poretti, 1986]{ant86} Antonello, E. \& Poretti, E., 1986, \aap, 169, 149
\bibitem[Antonello et al., 1987]{ant87} Antonello, E., Broglia, P., Conconi, P. \& Mantegazza, L., 1987, \aap, 171, 131
\bibitem[Antonello et al., 2000]{ant00} Antonello, E., Fugazza, D. \& Mantegazza, L., 2000, \aap, 356, L37
\bibitem[Buchler et al., 1990]{buc90} Buchler, R., Moskalik, P. \& Kov\'{a}cs, G., 1990, \apj, 351, 617
\bibitem[Buchler \& Moskalik, 1994]{buc94} Buchler, R. \& Moskalik, P., 1994, \aap, 292, 450
\bibitem[Dolphin et al., 2001]{dol01} Dolphin, A. et al., 2001, \apj, 550, 554
\bibitem[Dolphin et al., 2002]{dol02} Dolphin, A. et al., 2002, \aj, 123, 3154
\bibitem[Feast \& Walker, 1987]{fea87} Feast, M. \& Walker, A., 1987, \araa, 25, 345
\bibitem[Ferro et al., 1998]{fer98} Ferro, A. et al., 1998, \apjs, 117, 167
\bibitem[Freedman, 1988]{fre88} Freedman, W., 1988, \apj,  326, 691
\bibitem[Freedman et al., 1994]{fre94} Freedman, W. et al., 1994, \apj, 427, 628 
\bibitem[Freedman et al., 2001]{fre01} Freedman, W. et al., 2001, \apj,  553, 47 (H0KP)
\bibitem[Hendry et al., 1999]{hen99} Hendry, M., Tanvir, N. \& Kanbur, S., 1999, in \textit{Harmonizing Cosmic Distance Scales in a Post-HIPPARCOS Era}, ASP Conf. Series 167, eds. Egret \& Heck, Pg 192
\bibitem[Hintz \& Joner, 1997]{hin97} Hintz, E. \& Joner, M., 1997, \pasp, 109, 639
\bibitem[Herrnstein et al., 1999]{her99} Herrnstein, J. et al., 1999, \nat, 400, 539 
\bibitem[Kanbur \& Nikolaev, 2001]{kan01} Kanbur, S. \& Nikolaev, S., 2001, BAAS \#104.02
\bibitem[Kanbur et al., 2002]{kan02} Kanbur, S. et al., 2002, \apj \  submitted
\bibitem[Kennicutt et al., 1995]{ken95} Kennicutt, R., Freedman, W. \& Mould, J., 1995, \aj,  110, 1476
\bibitem[Madore \& Freedman, 1985]{mad85} Madore, B. \& Freedman, W., 1985, \aj, 90, 1104
\bibitem[Madore \& Freedman, 1991]{mad91} Madore, B. \& Freedman, W., 1991, \pasp, 103, 933
\bibitem[Moffett \& Barnes, 1985]{mof85} Moffett, T. \& Barnes, T.,  1985, \apjs, 58, 843
\bibitem[Moffett et al., 1998]{mof98} Moffett, T. et al., 1998, \apjs, 117, 135
\bibitem[Newman et al., 2001]{new01} Newman, J. et al., 2001, \apj, 553, 562
\bibitem[Paczy\'{n}ski \& Pindor, 2000] {pac00} Paczy\'{n}ski, B. \& Pindor, B., 2000, \apj, 533, L105
\bibitem[Poretti, 1994]{por94} Poretti, E., 1994, \aap, 285, 524
\bibitem[Press et al., 1992]{pre92} Press, W. et al, 1992, \textit{Numerical Recipes in C}, Cambridge University Press, $2^{nd}$ ed.
\bibitem[Saha \& Hoessel, 1990]{sah90} Saha, A. \& Hoessel, J., 1990, \aj , 99, 97
\bibitem[Saha et al., 1999]{sah99} Saha, A. et al., 1999, \apj, 522, 802
\bibitem[Saha et al., 2001a]{sah01a} Saha, A. et al., 2001a, \apj, 551, 973
\bibitem[Saha et al., 2001b]{sah01b} Saha, A. et al., 2001b, \apj, 562, 314
\bibitem[Saha et al., 2000c]{sah00c} Saha, A., Labhardt, L. \& Prosser, C., 2000c, \pasp, 112, 163 
\bibitem[Schaltenbrand \& Tammann, 1971]{sch71} Schaltenbrand, R. \& Tammann, G., 1971, \aap Suppl., 4, 265
\bibitem[Silbermann et al., 1996]{sil96} Silbermann, N. et al., 1996, \apj,  470, 1
\bibitem[Simon \& Lee, 1981]{sim81} Simon, N. \& Lee, A., 1981, \apj,  248, 291
\bibitem[Simon \& Davis, 1983]{sim83} Simon, N. \& Davis, C., 1983, \apj, 266, 787
\bibitem[Simon \& Moffett, 1985]{sim85} Simon, N. \& Moffett, T., 1985, \pasp, 97, 1078
\bibitem[Stetson, 1996]{ste96} Stetson, P., 1996, \pasp,  108, 851 
\bibitem[Tammann et al., 2001]{tam01} Tammann, G. et al., 2001, astro-ph 0112489
\bibitem[Tammann \& Reindl, 2002]{tam02} Tammann, G. \& Reindl, B., 2002, astro-ph 0208176
\bibitem[Tanvir, 1997]{tan97} Tanvir, N., 1997, in \textit{The Extra-galactic Distance Scale}, ST ScI May Symposium, eds. Livio, Donahue \& Panagia, Cambridge University Press, Pg 91
\bibitem[Tanvir et al., 1999]{tan99} Tanvir, N., Ferguson, H. \& Shanks, T., 1999, \mnras, 310, 175
\bibitem[Udalski et al., 1999a]{uda99a} Udalski, A. et al., 1999a, AcA, 49, 233
\bibitem[Udalski et al., 1999b]{uda99b} Udalski, A. et al., 1999b, ACA, 49, 437
\bibitem[van Genderen, 1978]{van78} van Genderen, A., 1978, \aap, 65, 147
\end{thebibliography}
\end{document}